\documentstyle[12pt,epsf]{article}

\begin{document}
\begin{center}{\Large
Self-Adjointness and Polarization of the Fermionic Vacuum in 
the Background of Nontrivial Topology\footnote{Talk presented at the
International Workshop "Mathematical Physics--today, Priority Technologies--for
tomorrow" (in honour of the 70th anniversary of Professor Walter Thirring's
birth), 12-17 May, Kyiv, Ukraine}}
\end{center}
\begin{center}
{\large Yu.A.~Sitenko\footnote{E-mail address: yusitenko@gluk.apc.org},
\qquad D.G.~Rakityansky\footnote{E-mail address: radamir@ap3.gluk.apc.org}}\\
\medskip
{\it Bogolyubov Institute for Theoretical Physics, \\
National Academy of Sciences of Ukraine,}\\
{\it  252143 Kiev, Ukraine}
\end{center}
\begin{abstract}
Singular configuration of an external static magnetic field
in the form of a string polarizes vacuum in the secondly 
quantized theory on a plane which is orthogonal to 
the string axis. We consider the most general boundary
conditions at the punctured singular point, which are 
compatible with the self-adjointness of the two-dimensional
Dirac Hamiltonian. The dependence of the induced vacuum 
quantum numbers on the self-adjoint extension parameter and
the flux of the string is determined. 
\end{abstract}

In this talk contributing to the celebration of the 70-th birthday of 
Professor Walter Thirring we would like to report on some exact results
concerning the properties of vacuum in the background of nonrivial topology.

As is known, the singular static magnetic monopole background induces
fermion number in the vacuum [1-3]
\begin{equation}
\langle N\rangle=-{1\over\pi}\arctan\left(\tan\frac{\Theta}{2}\right),
\end{equation}
where $\Theta$ is the parameter of a self-adjoint extension, which defines the
boundary condition at a puncture corresponding to the location of the monopole;
this results in the monopole becoming actually the dyon violating the Dirac
quantization condition and CP symmetry.

In the present talk we shall be considering quantum numbers which are induced
in the fermionic vacuum by the singular static magnetic string background.
Since
the deletion of a line , as compared to the deletion of a point, changes the
topology of space in a much more essential way (fundamental group becomes
nontrivial), the properties of the $\Theta$-vacuum will appear to be much more
diverse, as compared to Eq.(1).  Restricting ourselves to a surface which is
orthogonal to the string axis, let us consider 2+1-dimentional spinor
electrodynamics on a plane with a puncture corresponding to the location
of the string.  We shall show that in this case the induced vacuum fermion
number  and magnetic flux depend on the self-adjoint extension parameter
and the magnetic flux of the string as well.

The pertinent Dirac Hamiltonian has the form
\begin{equation}
H=-i\vec{\alpha}[\vec{\partial}-i\vec{V}(\vec{x})]+\beta m;
\end{equation}
where $\vec{V}(\vec{x})$ is an external static vector potential.  In a flat
two-dimensional space ($\vec{x}=(x^1,x^2)$) the vacuum fermion number induced
by such a background was calculated first in Ref.[4]
\begin{equation}
\langle N\rangle=-{1\over2}\Phi,
\end{equation}
where $m>0$ and $\Phi={1\over2\pi}\int d^2x B(\vec{x})$ is
the total flux (in
the units of $2\pi$) of the external magnetic field strength
$B(\vec{x})=\vec{\partial}\times \vec{V}(\vec{x})$ piercing the
two-dimensional space (plane).

It should be emphasized, however, that Eq.(3) is valid for  regular
external field configurations only, i.e. $B(\vec{x})= B_{\rm
reg}(\vec{x})$, where $B_{\rm reg}(\vec{x})$  is a
continuous in the whole function that can grow at most as
$O\bigl(|\vec{x}-\vec{x}_s|^{-2+\varepsilon}\bigr)$ $(\varepsilon>0)$
at separate points; as to a vector potential
$\vec{V}(\vec{x})=\bigl(V_1(\vec{x}),V_2(\vec{x})\bigr)$, it is
unambiguously defined everywhere on the plane. The regular
configuration of an external field polarizes the vacuum locally, and
Eq.(3) is just the integrated version of the linear relation between
the vacuum fermion number density and the magnetic field strength.

One can ask the following question: whether the nonlocal effects of
the external field background are possible, i.e., if the spatial
region of nonvanishing field strength is excluded, whether there will
be vacuum polarization in the remaining part of space? For the
positive answer it is necessary, although not sufficient, that the
latter spatial region be of nontrivial topology [5] (see
also Ref.[6]). However, the condition on the boundary of
the excluded region has not been completely specified. In the present
talk this point will be clarified by considering the whole set of
boundary conditions which are compatible with the self-adjointness
of the Dirac Hamiltonian in the remaining region.

We shall be interested in the situation when the volume of the
excluded region is shrinked to zero, while the global characteristics
of the external field in the excluded region is retained nonvanishing.
This implies that singular, as well as regular, configurations of
external fields have to be considered. In particular, in two spatial
dimensions the magnetic field strength is taken to be a distribution
(generalized function)
\begin{equation}
B(\vec{x})=B_{\rm reg}(\vec{x})+2\pi\Phi^{(0)}\delta(\vec{x}),
\end{equation}
where $\Phi^{(0)}$ is the total magnetic flux (in the units of $2\pi$)
in the excluded region which is placed at the origin $\vec{x}=0$. As
to the vector potential, it is unambiguously defined everywhere with
the exception of the origin, i.e. the limiting value
$\lim_{|\vec{x}|\to0}\vec{V}(\vec{x})$ does not exist, or, to be more
precise, a singular magnetic vortex is located at the origin
\begin{equation}
\lim_{|\vec{x}|\to0}\vec{x}\times\vec{V}(\vec{x})=\Phi^{(0)}.
\end{equation}
Certainly, a plane has trivial topology, $\pi_1=0$, while a plane with
a puncture where the vortex is located has nontrivial topology,
$\pi_1={{\rm Z}}$; here ${{\rm Z}}$ is the set of integer numbers and
$\pi_1$ is the first homotopy group of the surface.

Let us turn now to the boundary condition at the puncture $\vec{x}=0$.
In the following our concern will be in the case in which the regular part
of the magnetic field is absent, $B_{\rm reg}(\vec{x})=0$. Then, in
the representation with $\alpha_1=\sigma_1$,  $\alpha_2=\sigma_2$ and
$\beta=\sigma_3$ ($\sigma_j$ are the Pauli matrices) the spinor wave
function satisfying the Dirac equation has the
form
\begin{equation}
\psi(\vec{x})=\sum_{n\in{\rm Z}}\left(\begin{array}{l}f_n(r)\exp(in
\varphi)\\g_n(r)\exp[i(n+1)\varphi]\end{array}\right),
\end{equation}
where the radial functions, in general, are
\begin{equation}
\left(\begin{array}{l}f_n(r)\\g_n(r)\end{array}\right)=
\left(\begin{array}{c}C^{(1)}_n(E)J_{n-\Phi^{(0)}}
(kr)+C^{(2)}_n(E)Y_{n-\Phi^{(0)}}(kr)\\
{ik\over E+m}\bigl[C^{(1)}_n(E)J_{n+1-\Phi^{(0)}}
(kr)+C^{(2)}_n(E)Y_{n+1-\Phi^{(0)}}(kr)\bigr]\end{array}\right),
\end{equation}
$k=\sqrt{E^2-m^2}$, $J_\mu(z)$ and $Y_\mu(z)$ are the Bessel and the
Neumann functions of the order $\mu$. It is clear that the condition
of regularity at $r=0$ can be imposed on both $f_n$ and $g_n$ for all
$n$ in the case of integer values of the quantity
$\Phi^{(0)}$ only. Otherwise, the condition of regularity at
$r=0$ can be imposed on both $f_n$ and $g_n$ for all but $n=n_0$,
where
\begin{equation}
n_0={[\![}\Phi^{(0)}{]\!]}\;,
\end{equation}
${[\![} u{]\!]}$ is the integer part of the quantity $u$ (i.e. the
integer which is less than or equal to $u$); in this case at least one
of the functions, $f_{n_0}$ or $g_{n_0}$, remains irregular, although
square integrable, with the asymptotics $r^{-p}$ $(p<1)$ at $r\to0$.
The question arises then, what boundary condition, instead of
regularity, is to be imposed on $f_{n_0}$ and $g_{n_0}$ at $r=0$ in
the latter case?

To answer this question, one has to find the self-adjoint extension
for the partial Hamiltonian  corresponding to the mode with $n=n_0$.
If this Hamiltonian is defined on the domain of regular at $r=0$
functions, then it is Hermitian, but not self-adjoint, having the
deficiency index equal to (1,1). Hence the family of self-adjoint
extensions is labeled by one real continuous parameter [7] denoted in the
following by $\Theta$. It can be shown (see Ref.[8]) that,
for the partial Hamiltonian to be self-adjoint, it has to be defined
on the domain of functions satisfying the boundary condition 
\begin{equation}
\lim_{r\to0}\cos\biggl({\Theta\over2}+{\pi\over4}\biggr)\biggl
(mr\biggr)^Ff_{n_0}(r)=i\lim_{r\to0}\sin\biggl({\Theta\over2}+{\pi\over4}
\biggr)\biggl(mr\biggr)^{1-F}g_{n_0}(r),
\end{equation}
where
\begin{equation}
F={\{\hspace{-3.3pt}|}\Phi^{(0)}{\}\!\!\!|}\;,
\end{equation}
${\{\hspace{-3.3pt}|} u{\}\!\!\!|}$ is the fractional part of the quantity $u$,
${\{\hspace{-3.3pt}|} u{\}\!\!\!|}=u-{[\![} u{]\!]}$, $0\leq{\{\hspace{-3.3pt}|} u{\}\!\!\!|}<1$; note here that
Eq.(9) implies that $0<F<1$, since in the case of $F=0$ both
$f_{n_0}$ and $g_{n_0}$ satisfy the condition of regularity at $r=0$.

Using the explicit form of the solution to the Dirac equation in the
background of a singular magnetic vortex, it is straightforward to
calculate the vacuum fermion number induced on a punctured plane.
As follows already from the preceding discussion, the vacuum fermion
number vanishes in the case of integer values of $\Phi^{(0)}$
$(F=0)$, since this case is indistinguishable from the case of the trivial
background, $\Phi^{(0)}=0$. In the case of noninteger values
of $\Phi^{(0)}$ ($0<F<1$)
we get [9]
\begin{equation}
\langle N\rangle=\left\{
	\begin{array}{lc}
	\hphantom{-}\frac{1}{2}(1-F),&-1<A<\infty\\&\\
	-\frac{1}{2}(1+F),&-\infty<A<-1\\&\\
	-\frac{1}{2}F,&A^{-1}=0
	\end{array}\right\}, 
\qquad 0<F<\frac{1}{2} 
\end{equation}
\begin{equation}
\langle N\rangle=-\frac{1}{\pi}
	\arctan\biggl(\tan{\displaystyle\Theta\over2}\biggr),
\hspace*{4cm}F={1\over2}, \\%[5mm]
\end{equation}
\begin{equation}\label{F1}
\langle N\rangle=\left\{
	\begin{array}{lc}
	-\frac{1}{2}F,&-1<A^{-1}<\infty\\&\\
	\hphantom{-}\frac{1}{2}(2-F),&-\infty<A^{-1}<-1\\&\\
	\hphantom{-}\frac{1}{2}(1-F),&A=0
	\end{array}\right\}, 
		\frac{1}{2}<F<1.
\end{equation}
where
\begin{equation}
A=2^{1-2F}{\Gamma(1-F)\over\Gamma(F)}\tan
\biggl({\Theta\over2}+{\pi\over4}\biggr),
\end{equation}
$\Gamma(u)$ is the Euler gamma-function;
note that Eq.(12) coincides with Eq.(1).

It is obvious that the vacuum fermion number at fixed values of
$\Theta$ is periodic in the value of
$\Phi^{(0)}$. This feature (periodicity in $\Phi^{(0)}$) is also
shared by the quantum-mechanical scattering of a nonrelativistic
particle in the background of a singular magnetic vortex, known as the
Aharonov-Bohm effect [10].

The total magnetic flux induced in the fermionic vacuum on a punctured plane
is also calculated in Ref. [9]
\begin{equation}
\Phi^{(I)}=-{e^2F(1-F)\over2\pi m}\left[{1\over6}\biggl(F-{1\over2}\biggr)+
{1\over4\pi}\int\limits_1^\infty{dv\over
v\sqrt{v-1}}{Av^F-A^{-1}v^{1-F}\over Av^F+2+A^{-1}v^{1-F}}\right];
\end{equation}
note that the coupling constant $e$ relating the vacuum current to the
vacuum magnetic field strength (via the Maxwell equation) has the
dimension $\sqrt{m}$ in $2+1$-dimensional space-time. At
half-integer values of $\Phi^{(0)}$ we get
\begin{equation}
\Phi^{(I)}=-{e^2\over8\pi^2m}\arctan\biggl(\tan
{\Theta\over2}\biggr)
\qquad\left(F=\frac{1}{2}\right).
\end{equation}

Functional dependence of the vacuum fermion number and magnetic flux on
$\Theta$ in the range $-\frac{\pi}{2}<\Theta<\frac{3\pi}{2}$ at some fixed
values of $F$ is depicted in Figs. 1 and 2.
Functional dependence of the vacuum quantum numbers on
$F$ at some fixed values of $\Theta$ is depicted in Figs. 3 and 4.
Note that, depending on the choice of boundary condition, the vacuum can be
either of paramagnetic (sgn($\Phi^{(I)}$)=sgn($\Phi^{(0)}$)) or diamagnetic
(sgn($\Phi^{(I)}$)=-sgn($\Phi^{(0)}$)) type.

In conlusion we note that under the charge conjugation,
\begin{equation}
C:\quad \vec{V}\to-\vec{V},\quad \psi\to\sigma_1\psi^*,
\end{equation}
the fermion number operator and its vacuum value, as well as the vacuum
magnetic flux, are to be odd.
Evidently, the results
(11)--(13) and (15) are not, since the boundary condition (9) breaks,
in general, the
charge conjugation symmetry. However, for certain choices of the
parameter $\Theta$ this symmetry can be retained.  The most general boundary
condition which is compatible both with the periodicity in $\Phi^{(0)}$
and charge conjugation symmetry has the form
\begin{equation}
\Theta=\left\{
\begin{array}{lc}
\hphantom{-}\Theta_C({\rm mod}2\pi),&0<F<\frac{1}{2}\\&\\
\hphantom{-}0_{\hphantom{C}}({\rm mod}2\pi),       &F=\frac{1}{2}\\&\\
           -\Theta_C({\rm mod}2\pi),&\frac{1}{2}<F<1
\end{array}\right\}, -\pi<\Theta_C\leq\pi.
\end{equation}

%\vspace*{1.5cm}
The research was supported in part by the Ministry of Science and
Technologies of Ukraine.

\newpage
\begin{center}
\begin{tabular}{cc}%\hline
\epsfxsize=7cm\epsffile{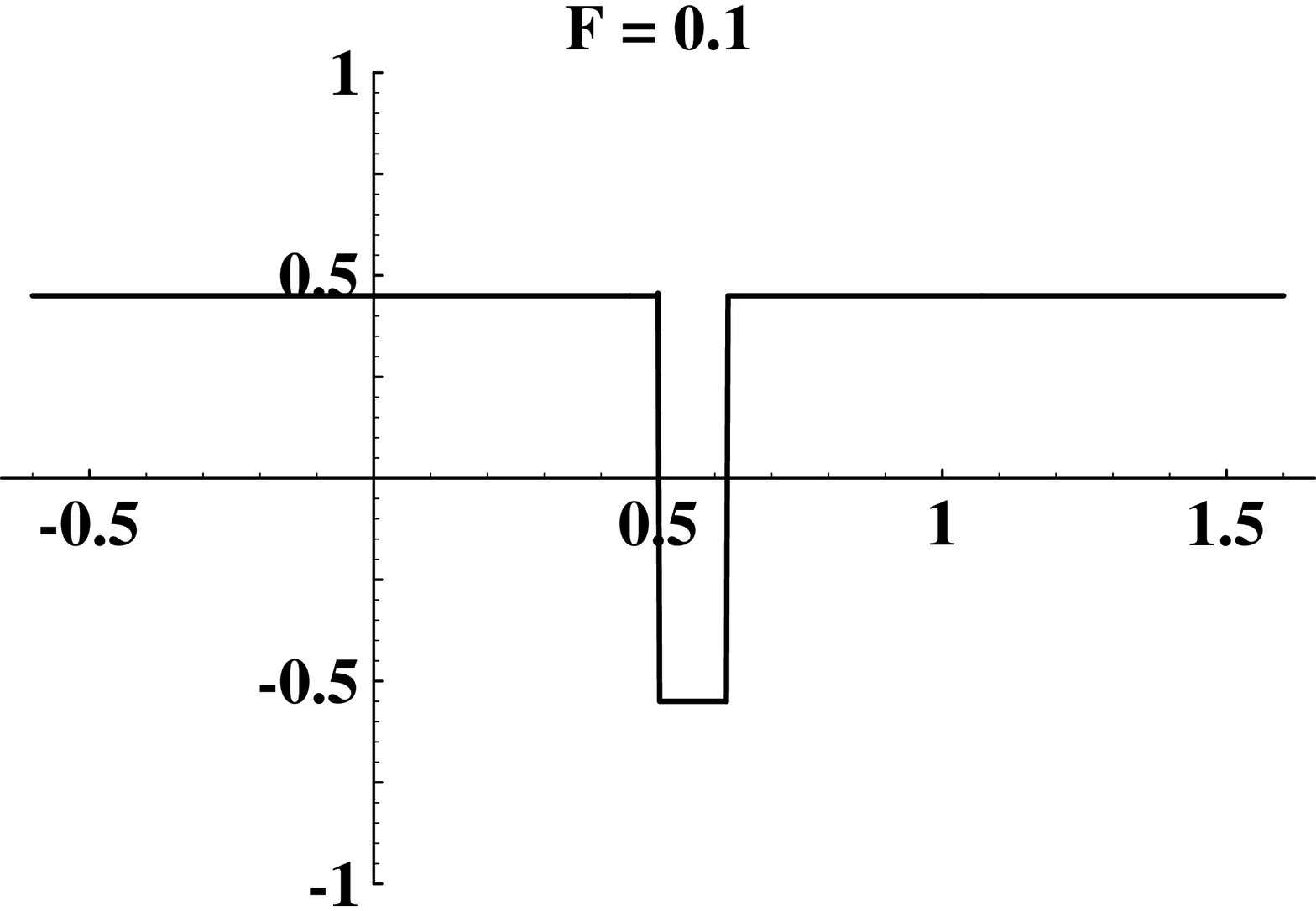}&\epsfxsize=7cm\epsffile{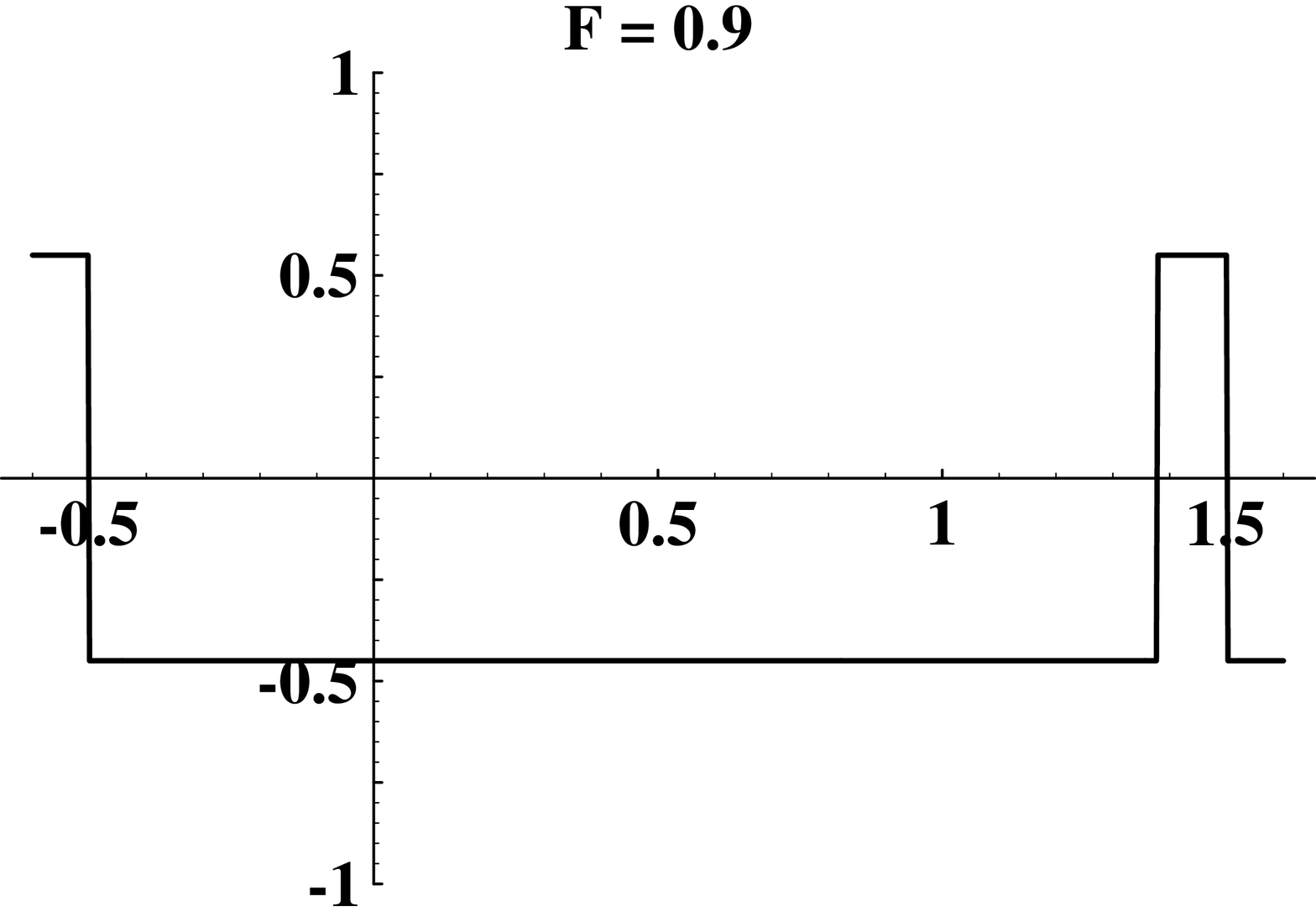}\\%\hline
\epsfxsize=7cm\epsffile{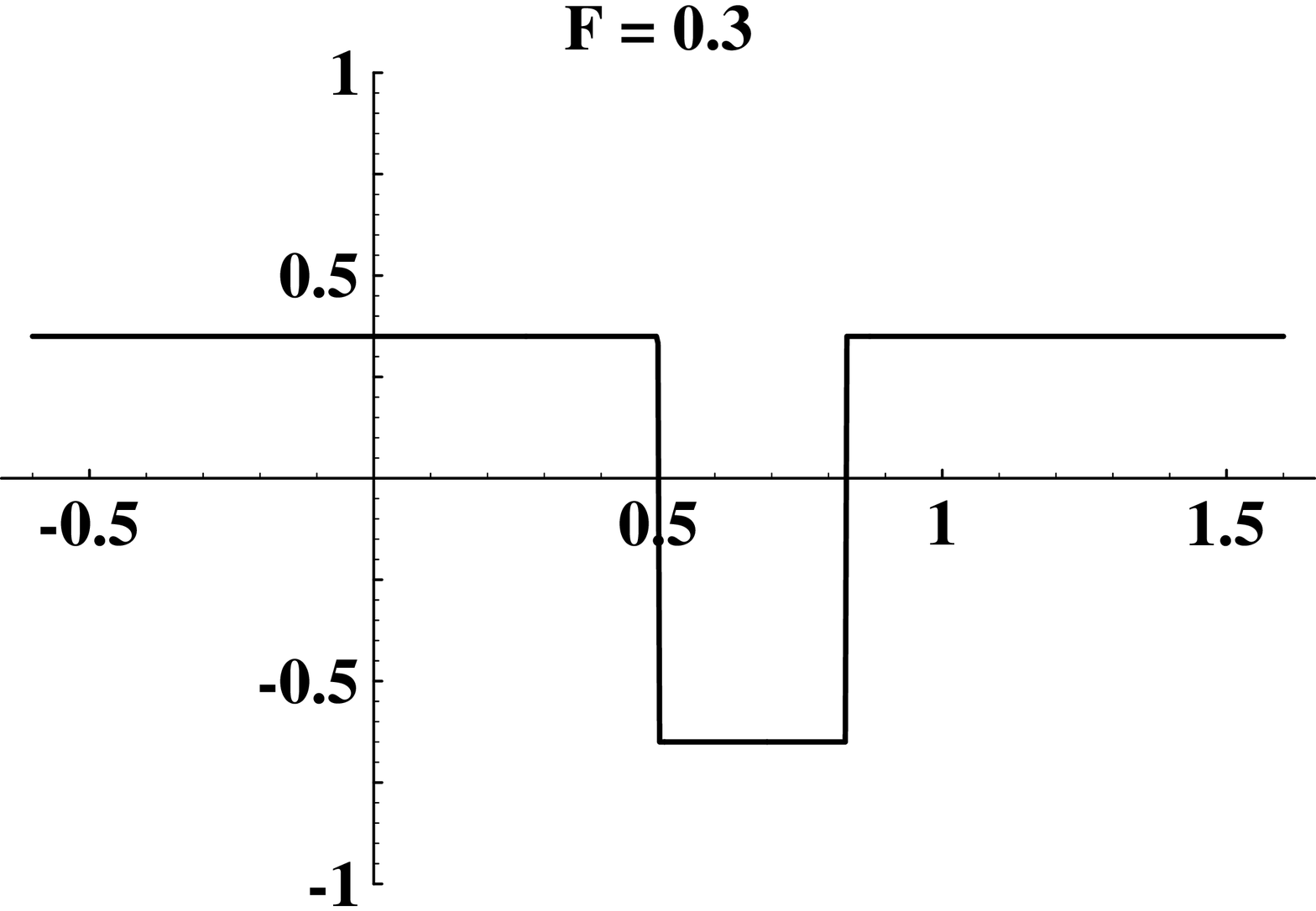}&\epsfxsize=7cm\epsffile{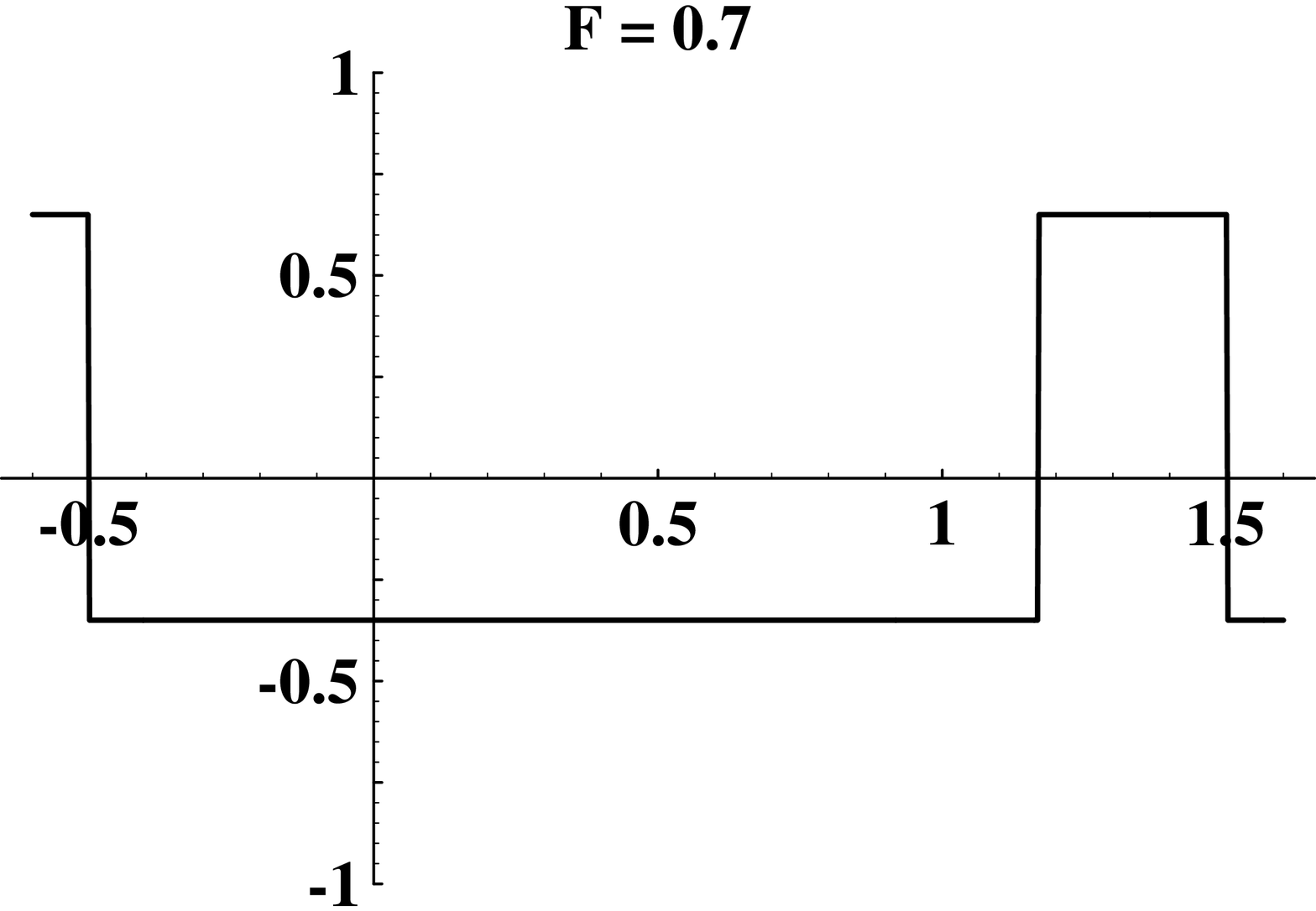}\\%\hline
\epsfxsize=7cm\epsffile{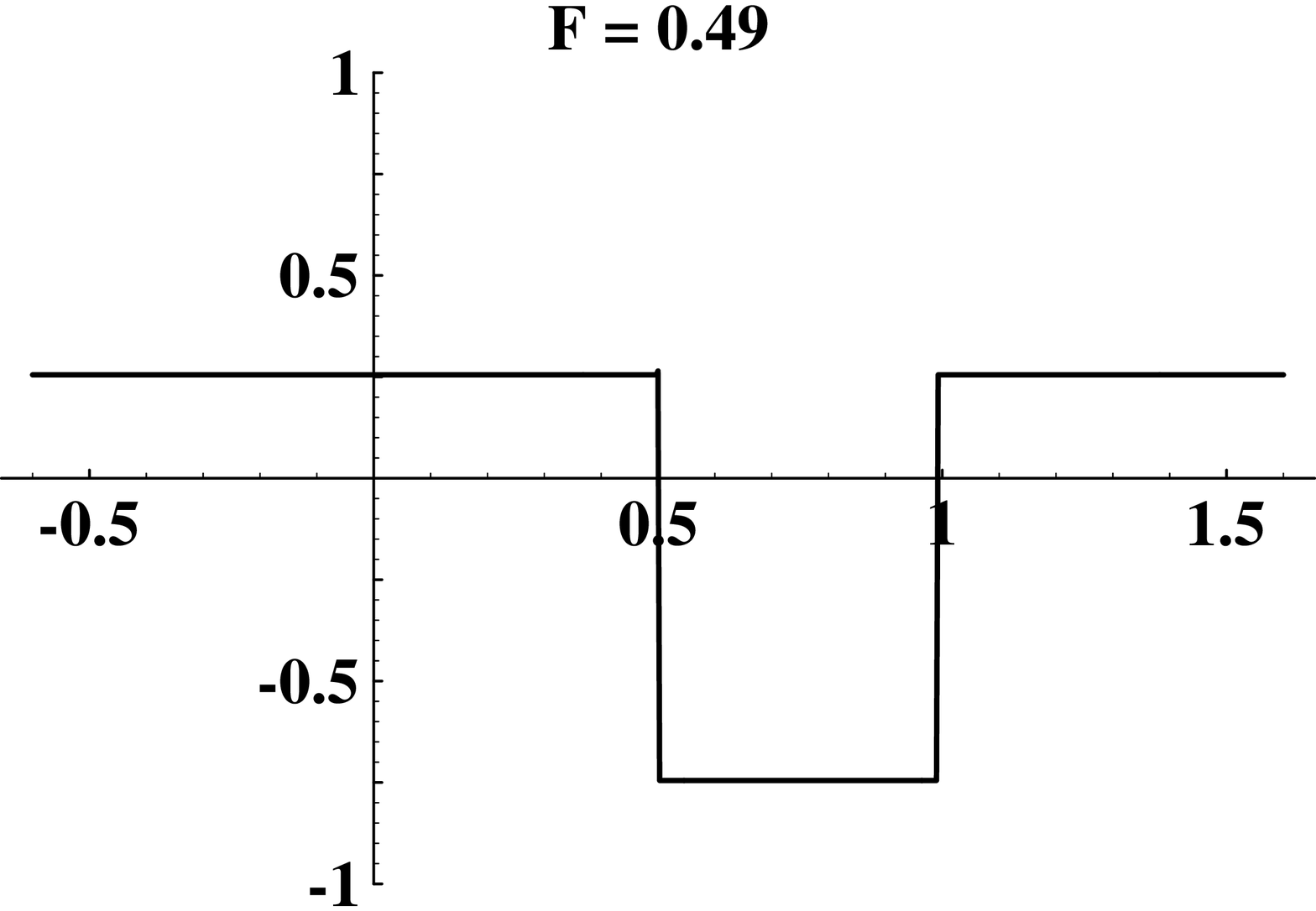}&\epsfxsize=7cm\epsffile{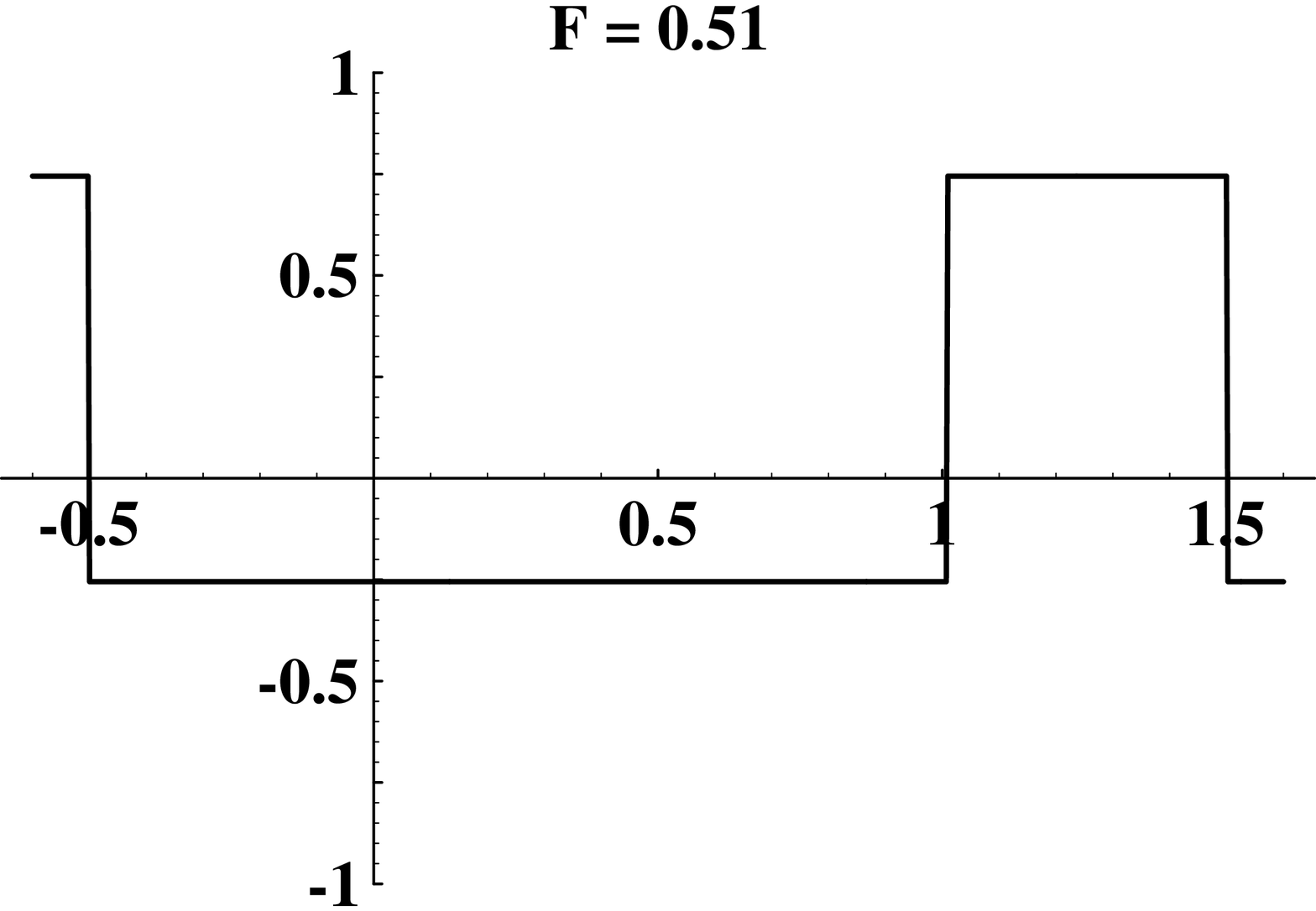}\\%\hline
\multicolumn{2}{c}{\epsfxsize=7cm\epsffile{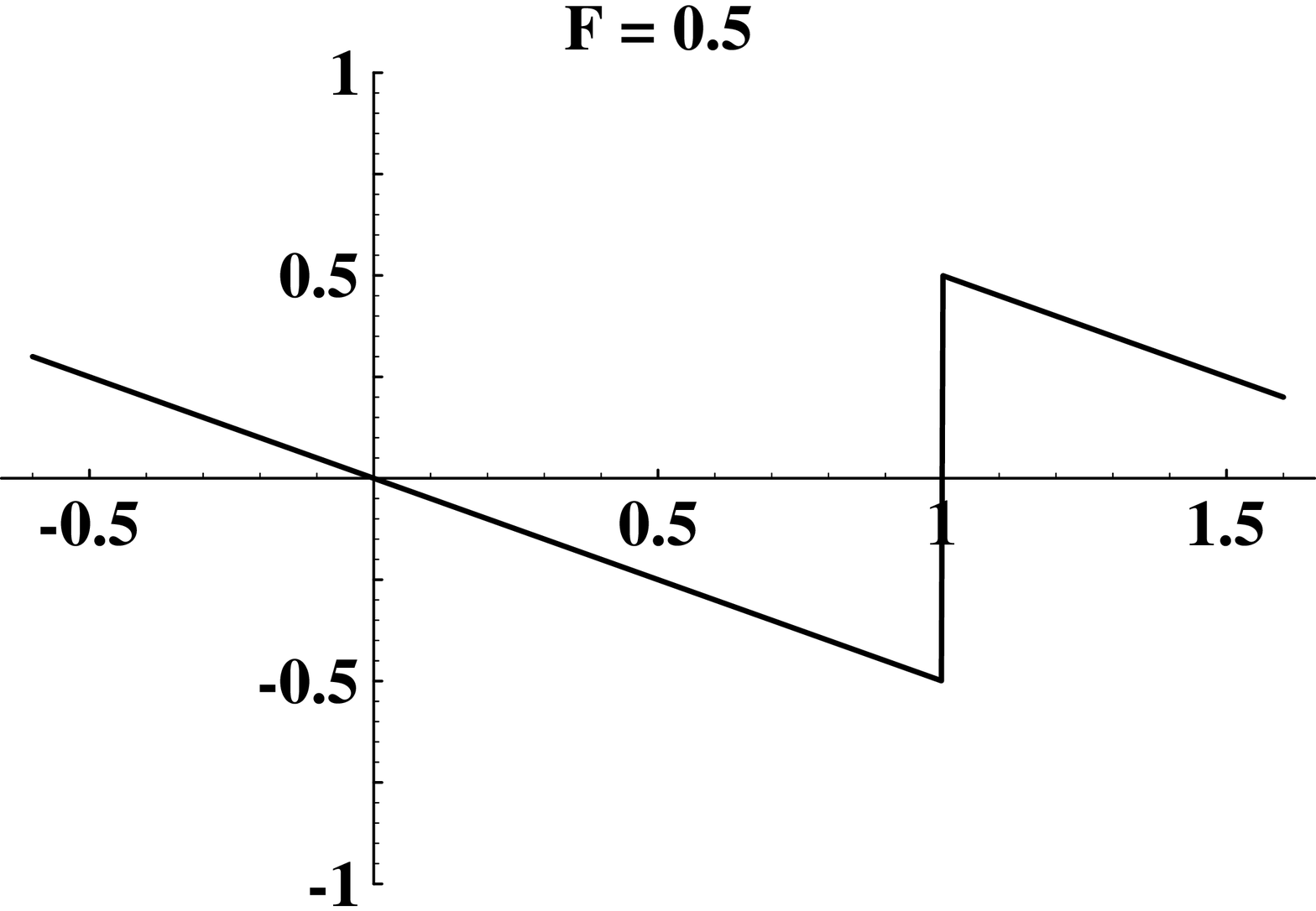}}\\%\hline
\end{tabular}
%\end{center}
\vfill
Fig. 1. $\langle N\rangle$ as function of $\Theta/\pi$. 
\newpage
%\begin{center}
\begin{tabular}{cc}%\hline
\epsfxsize=7cm\epsffile{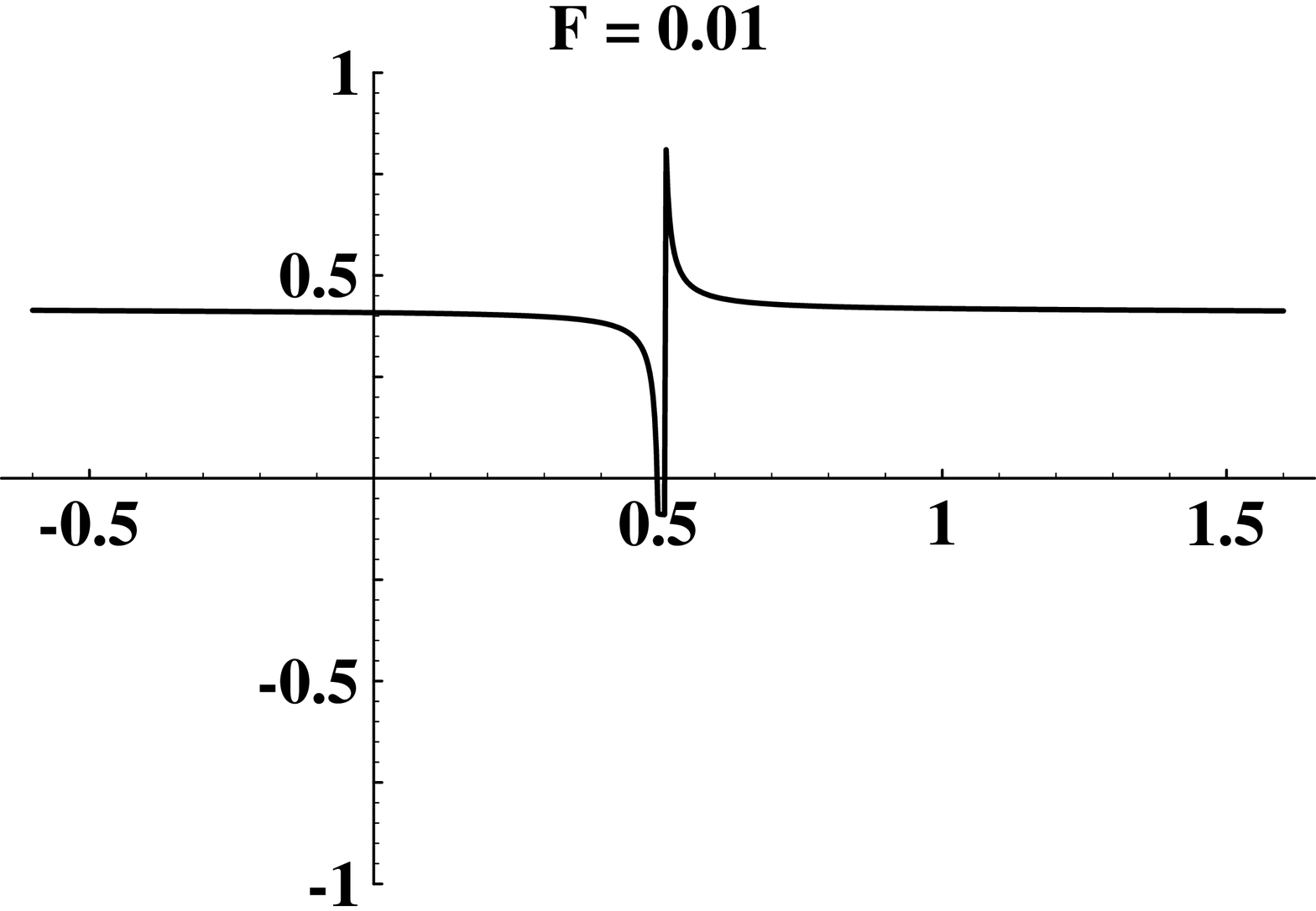}&\epsfxsize=7cm\epsffile{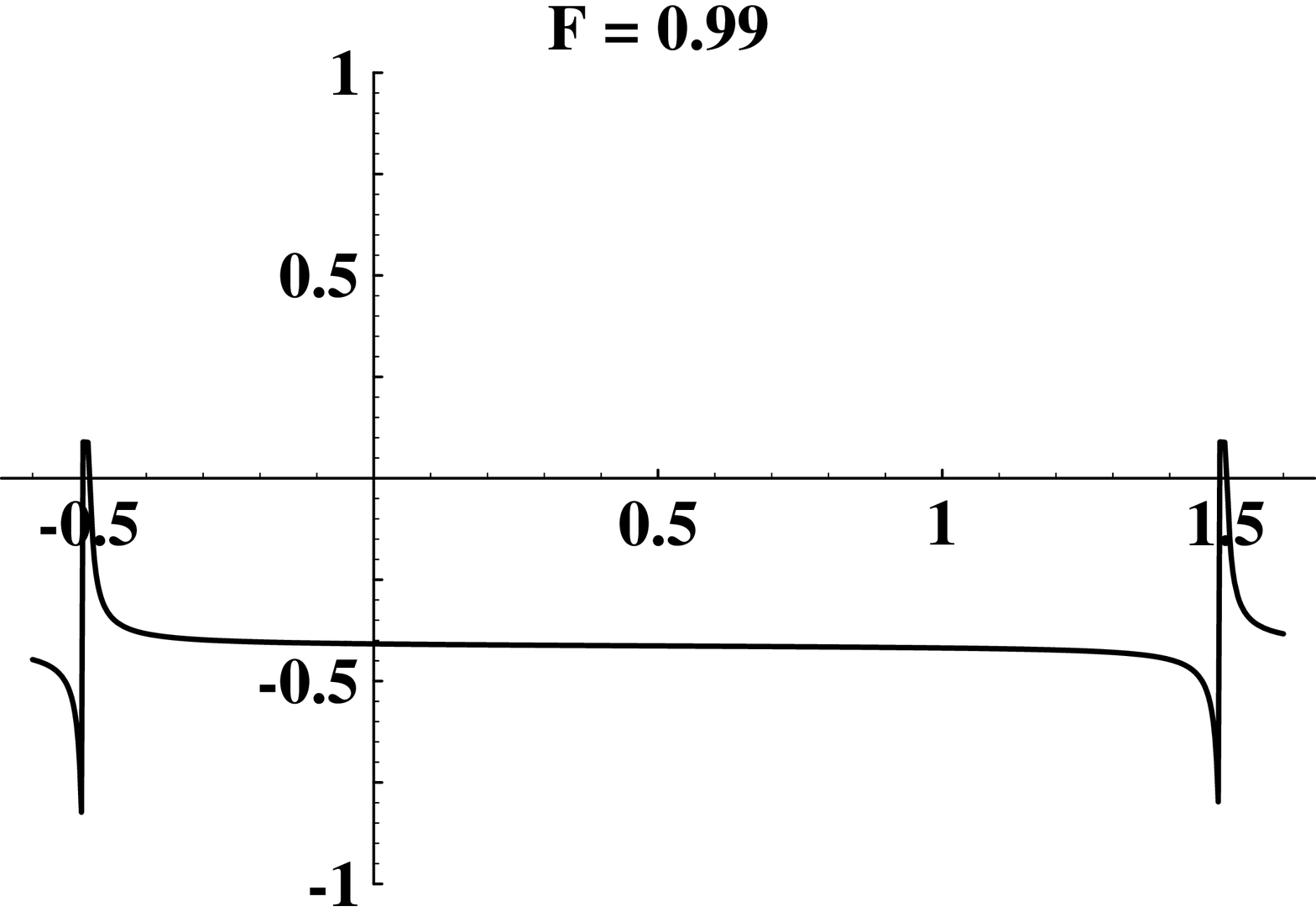}\\%\hline
\epsfxsize=7cm\epsffile{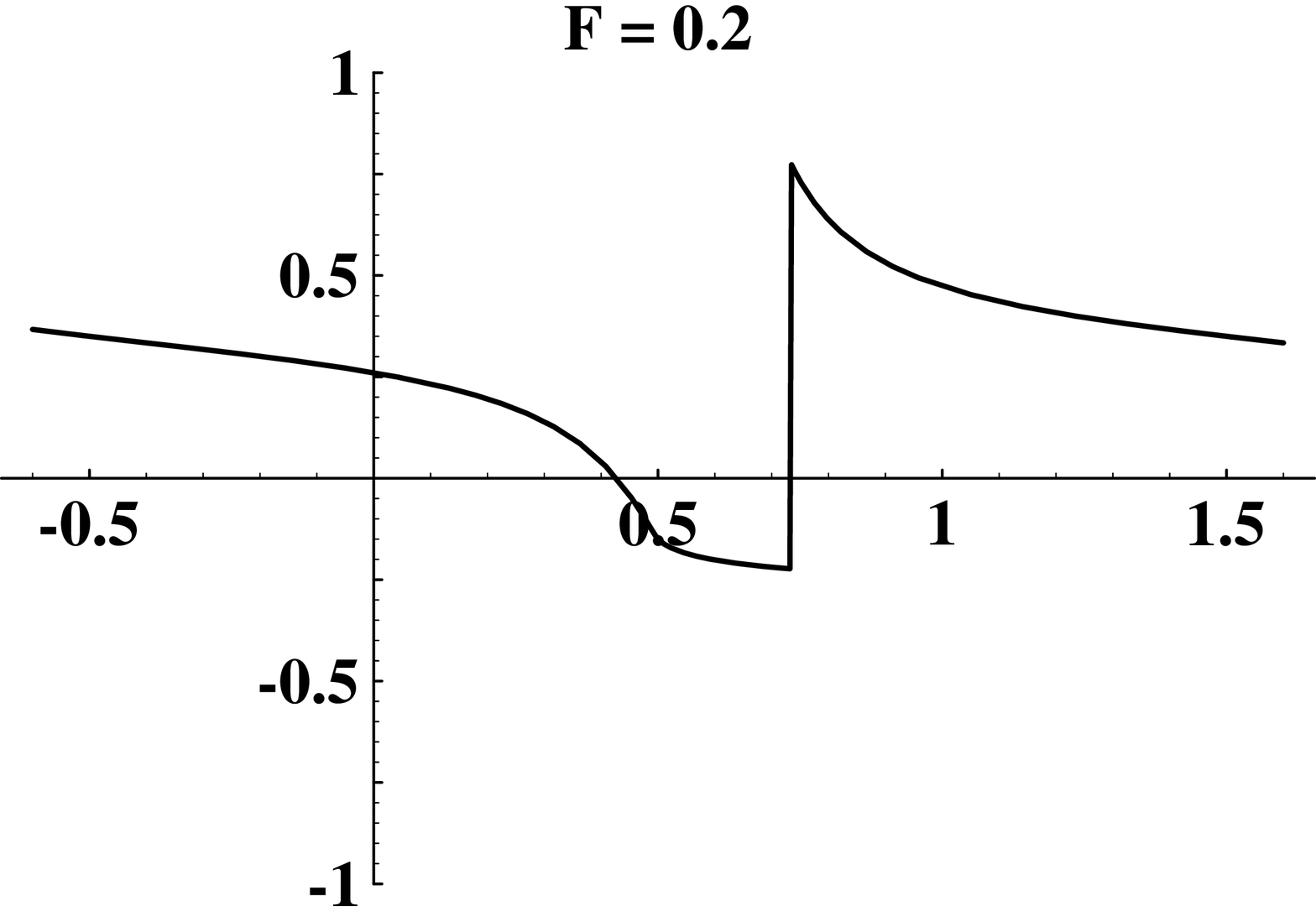}&\epsfxsize=7cm\epsffile{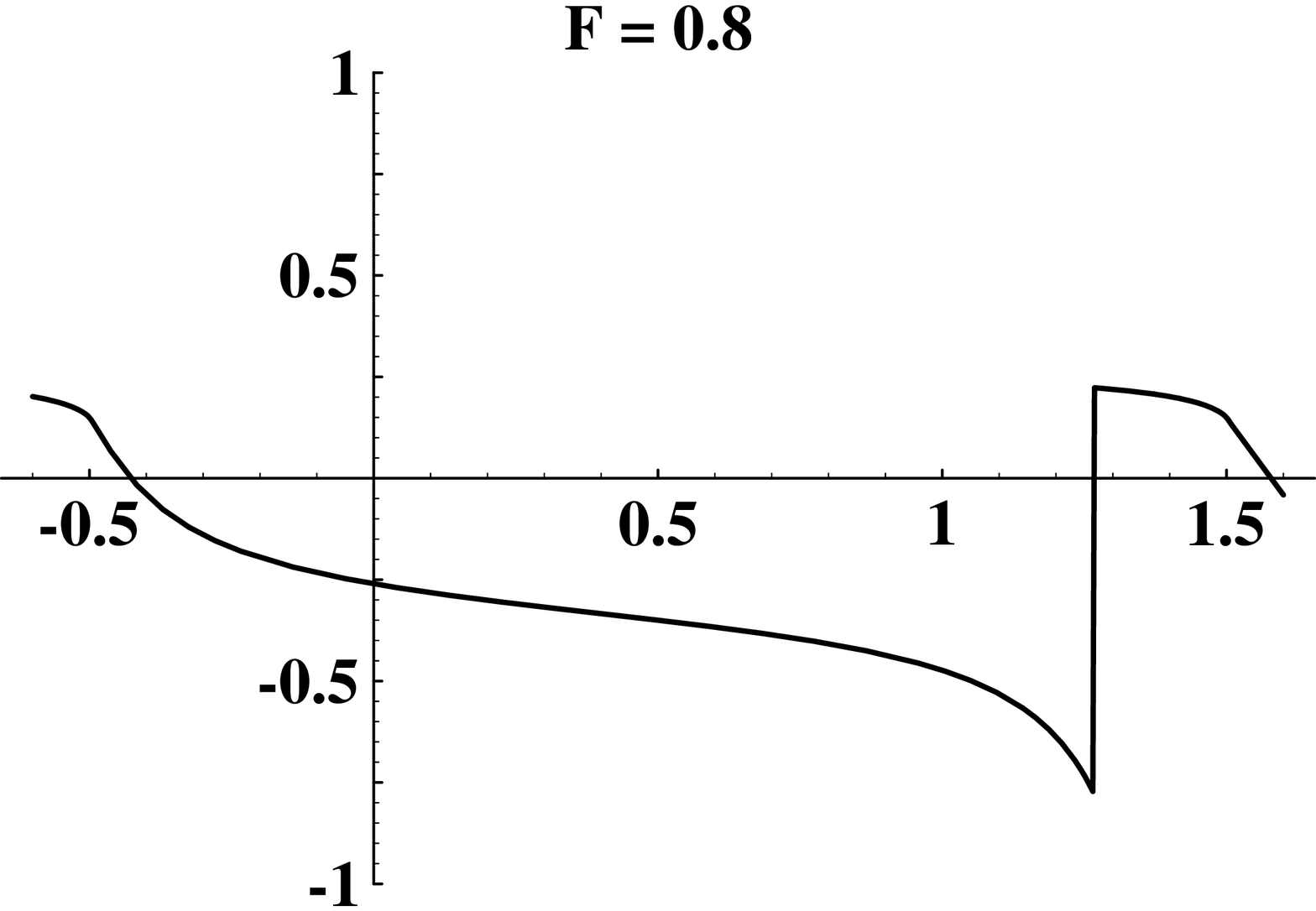}\\%\hline
\epsfxsize=7cm\epsffile{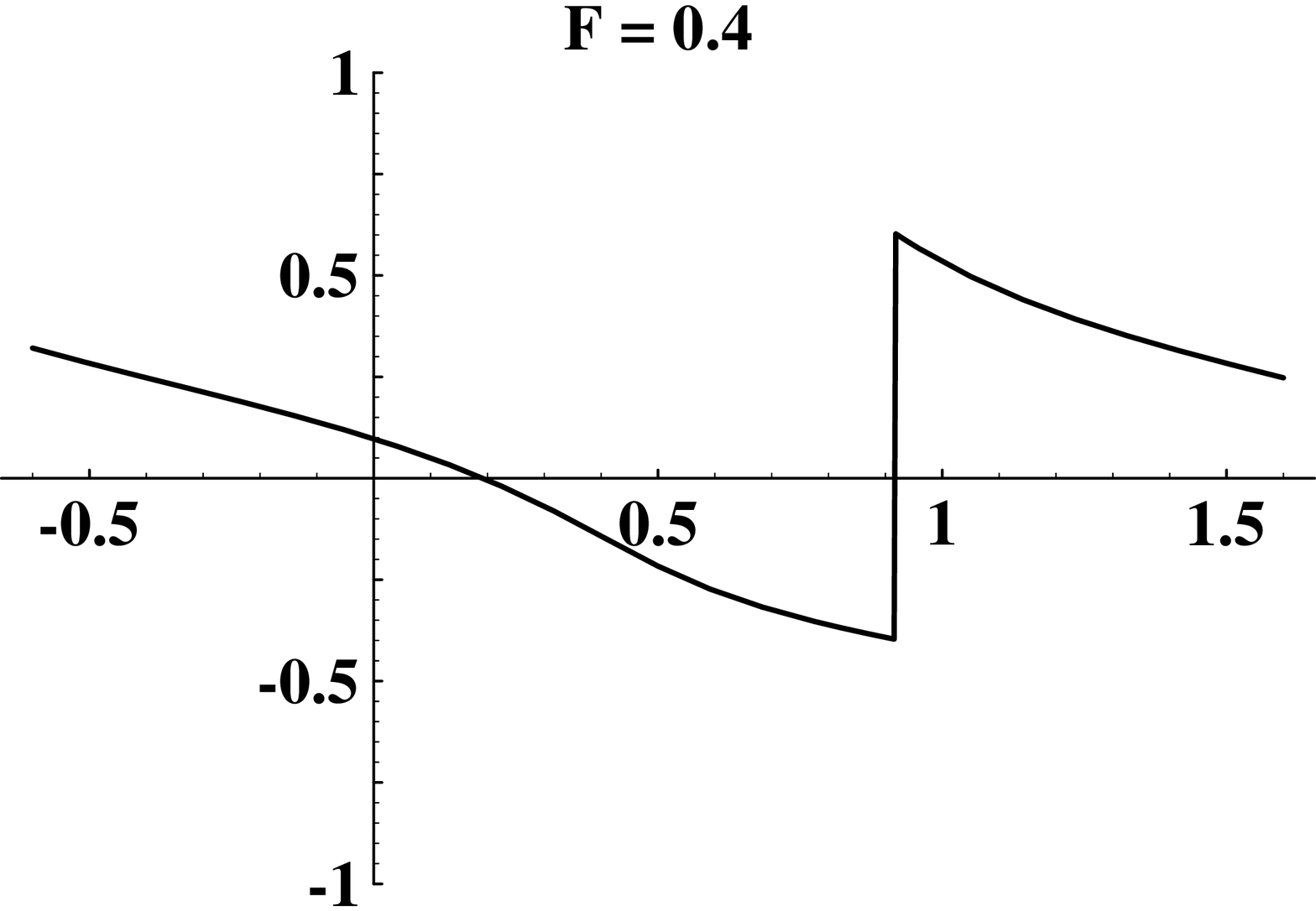}&\epsfxsize=7cm\epsffile{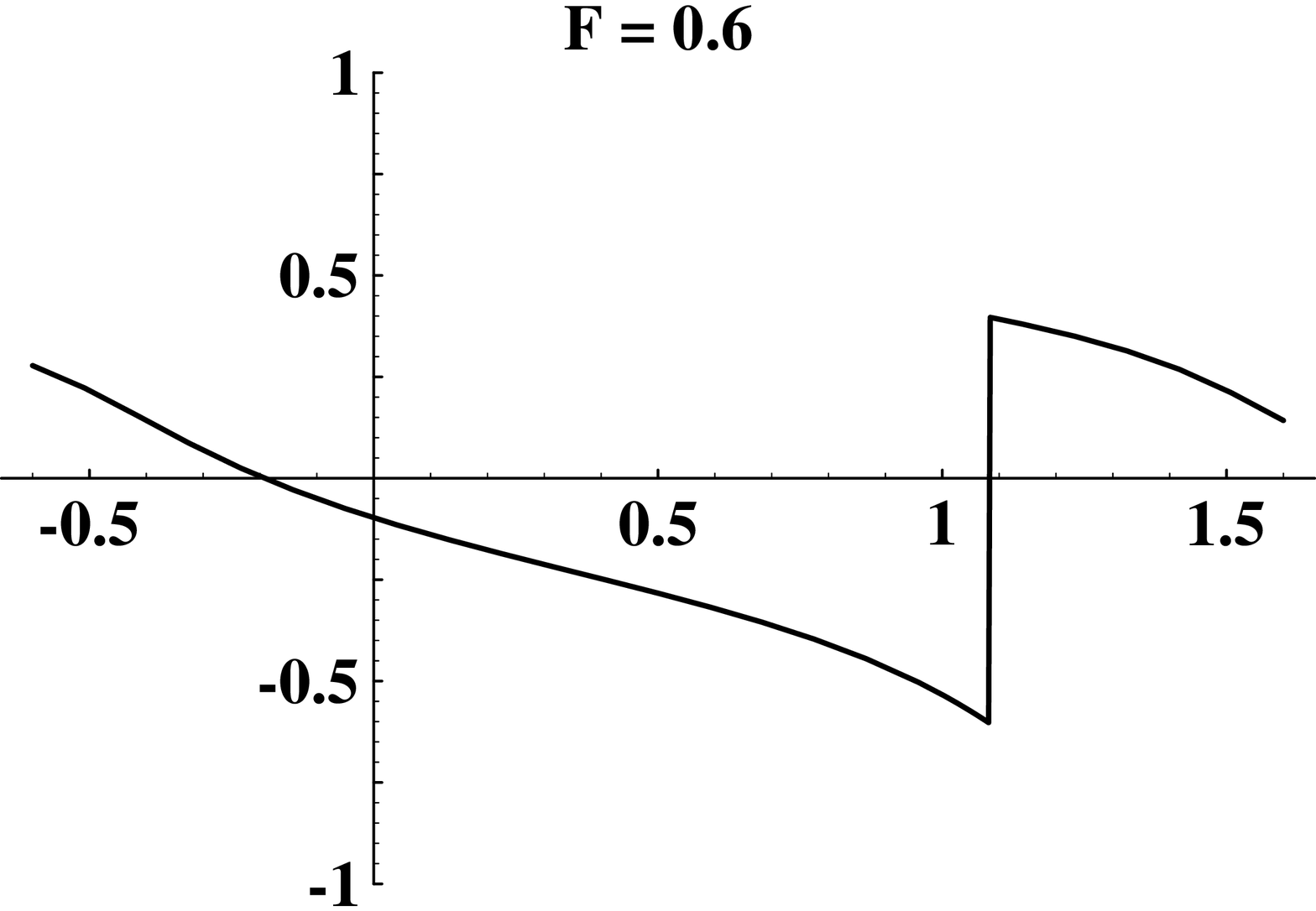}\\%\hline
\multicolumn{2}{c}{\epsfxsize=7cm\epsffile{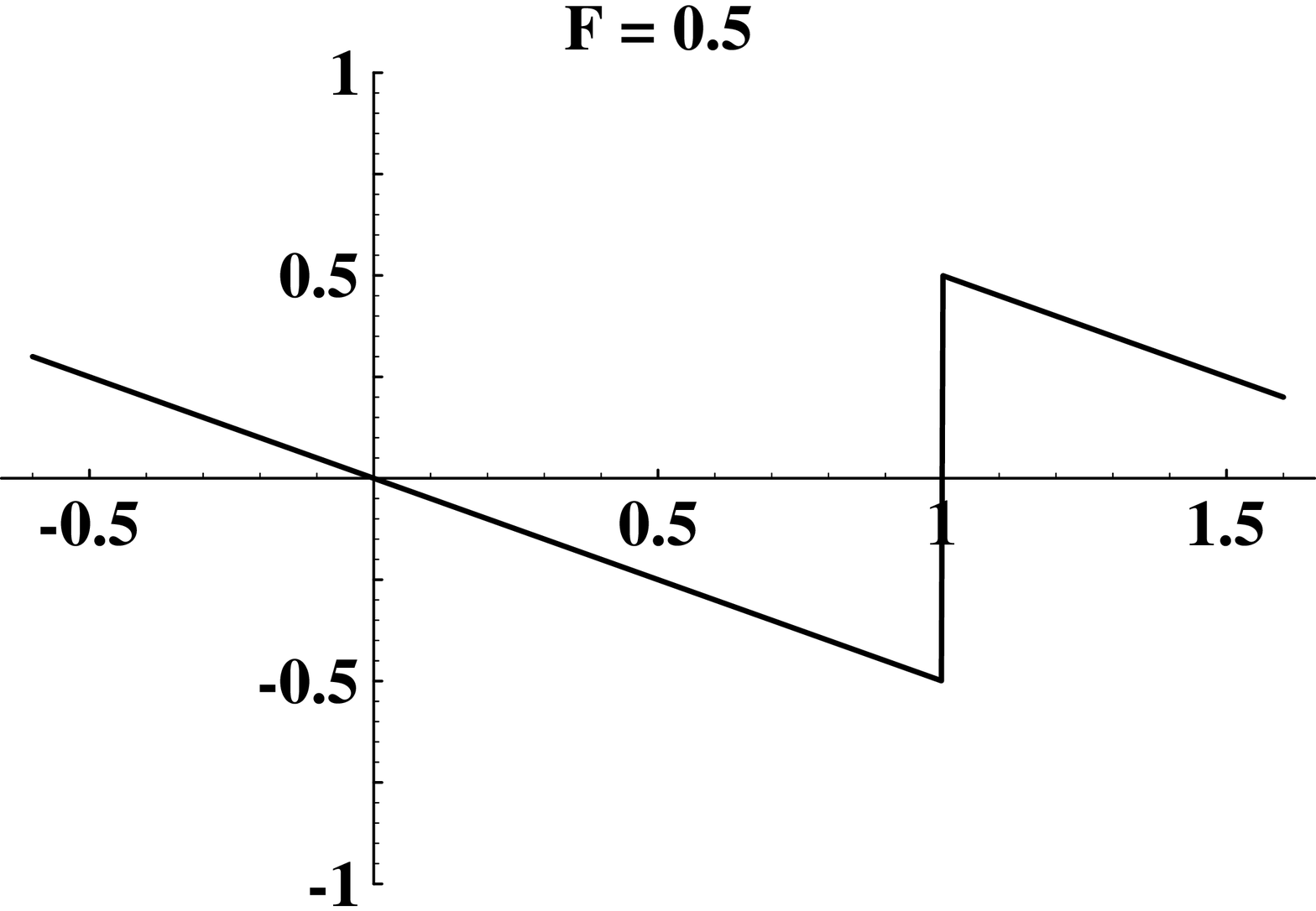}}\\%\hline
\end{tabular}
%\end{center}
\vfill
Fig. 2. $2\pi m[e^2F(1-F)]^{-1}\Phi^{(I)}$ as function of $\Theta/\pi$.
\newpage
%\begin{center}
\begin{tabular}{cc}%\hline
\epsfxsize=7cm\epsffile{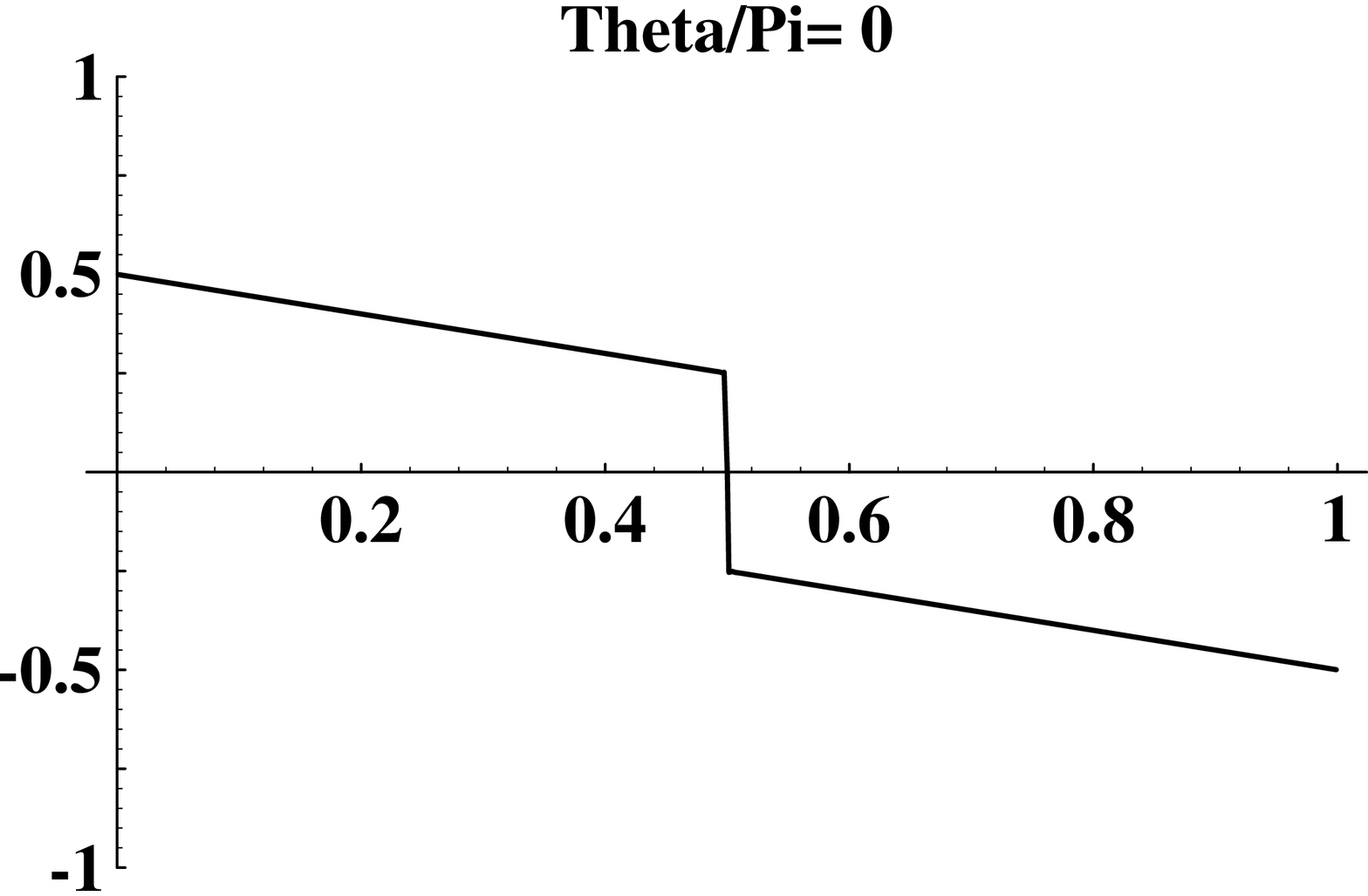}&\epsfxsize=7cm\epsffile{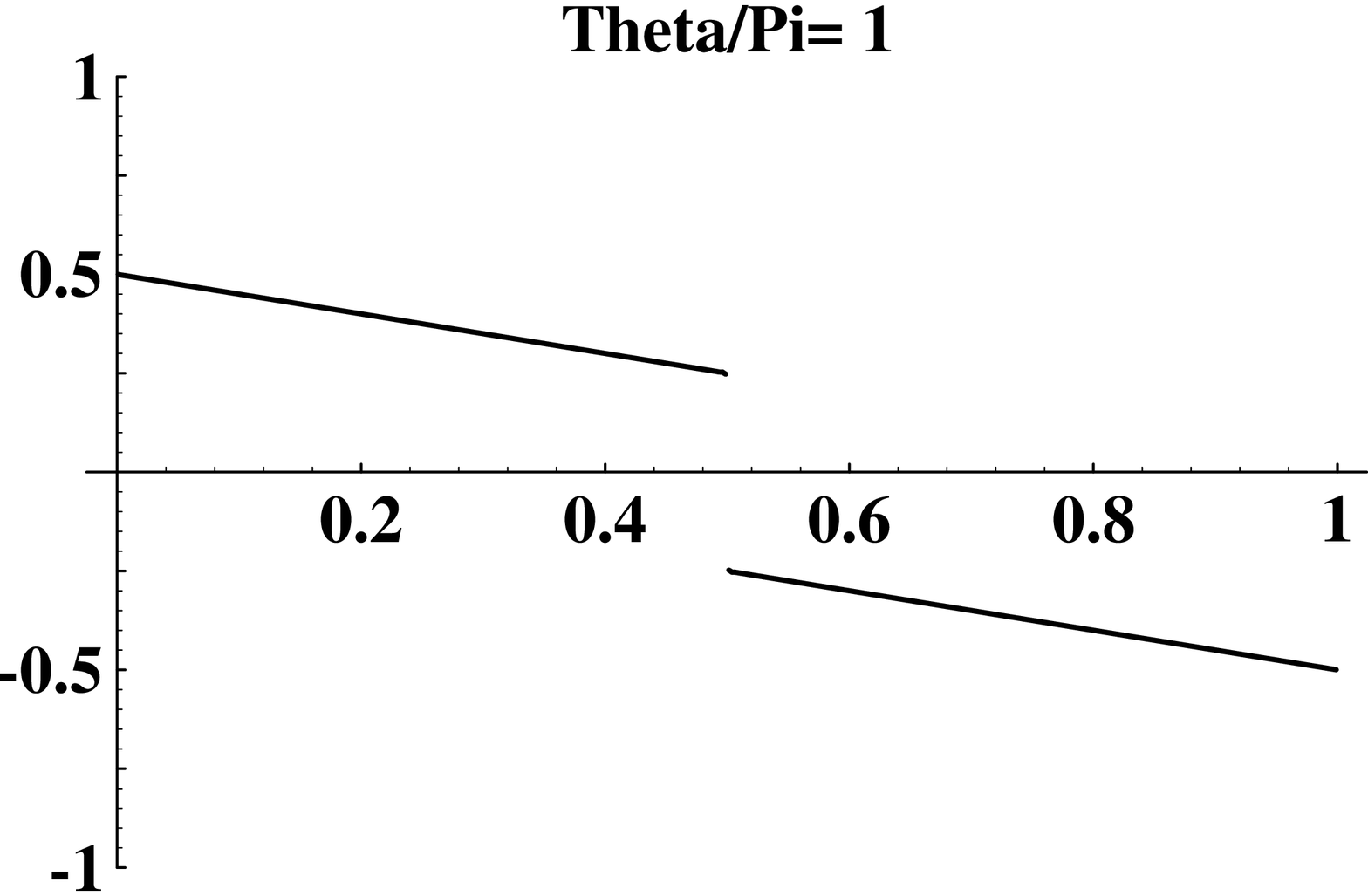}\\%\hline
\epsfxsize=7cm\epsffile{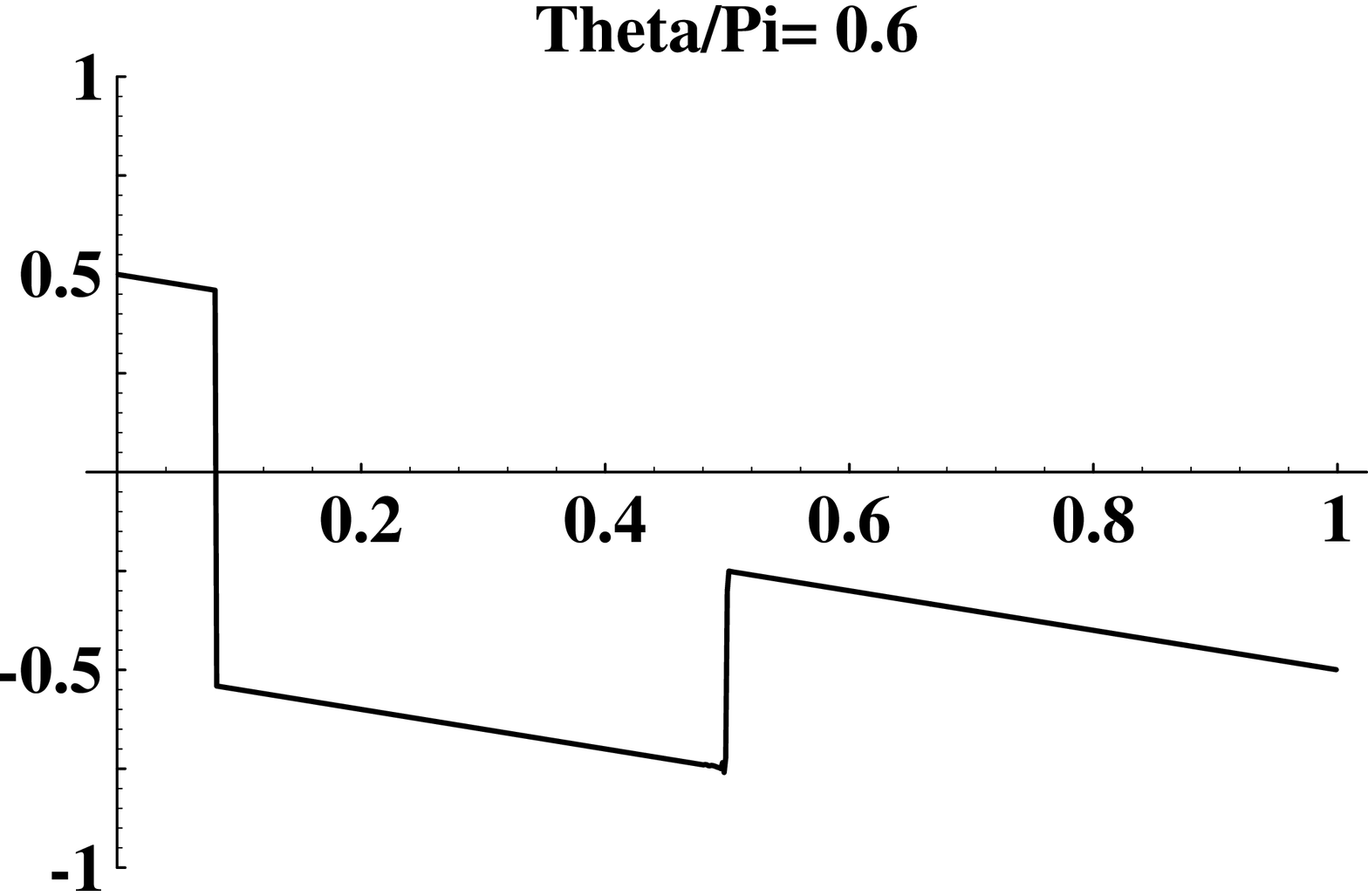}&\epsfxsize=7cm\epsffile{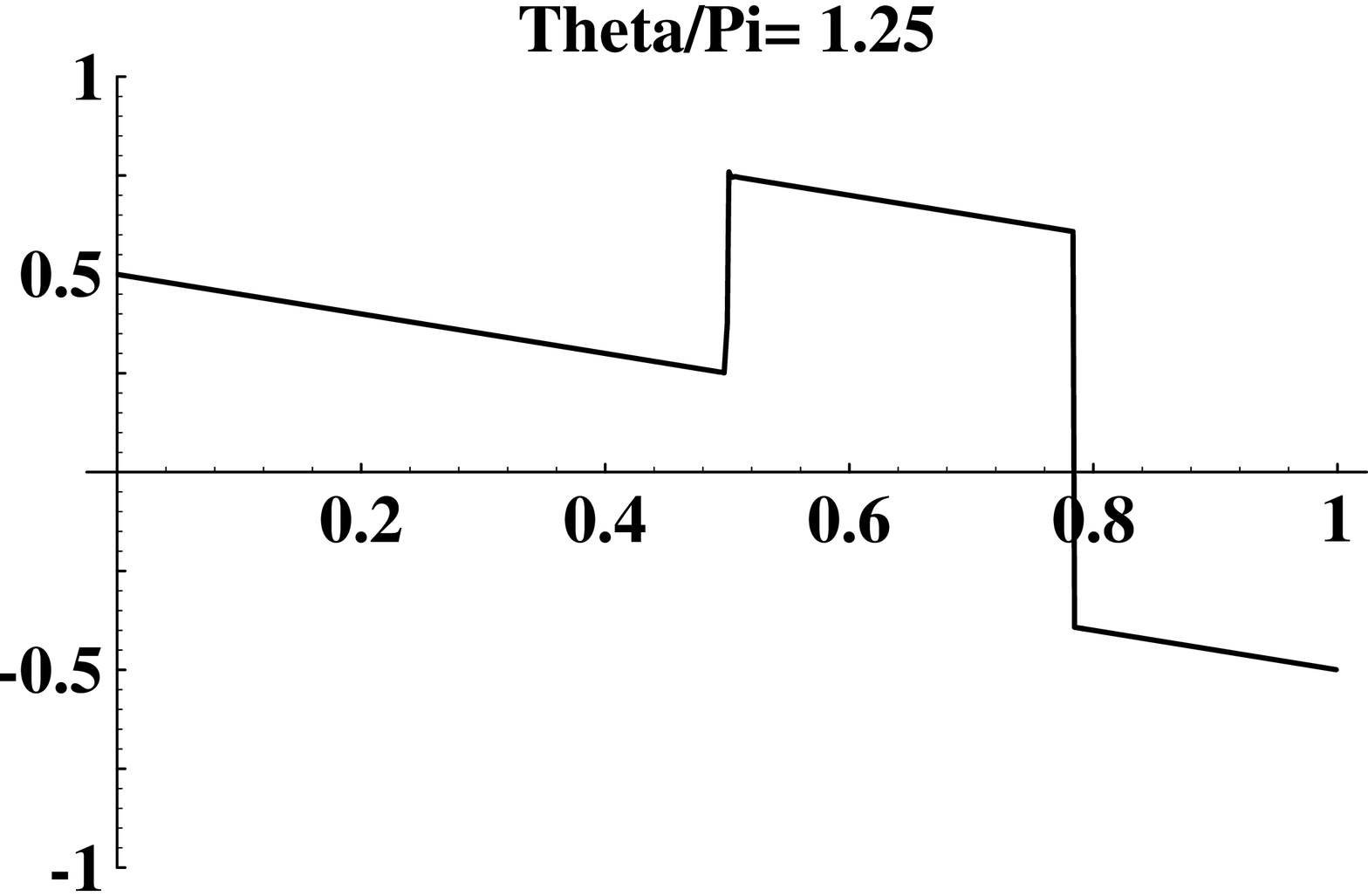}\\%\hline
\epsfxsize=7cm\epsffile{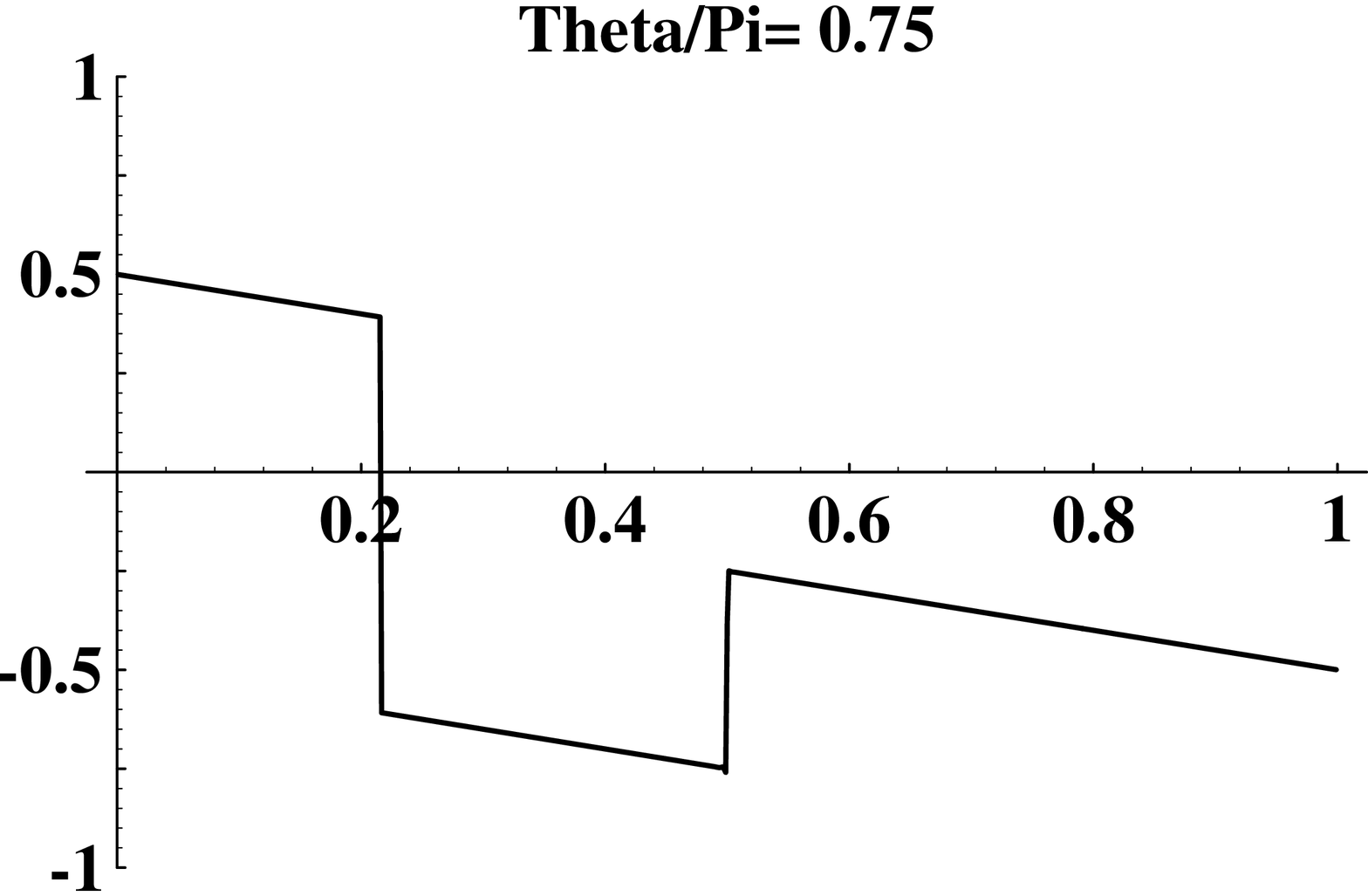}&\epsfxsize=7cm\epsffile{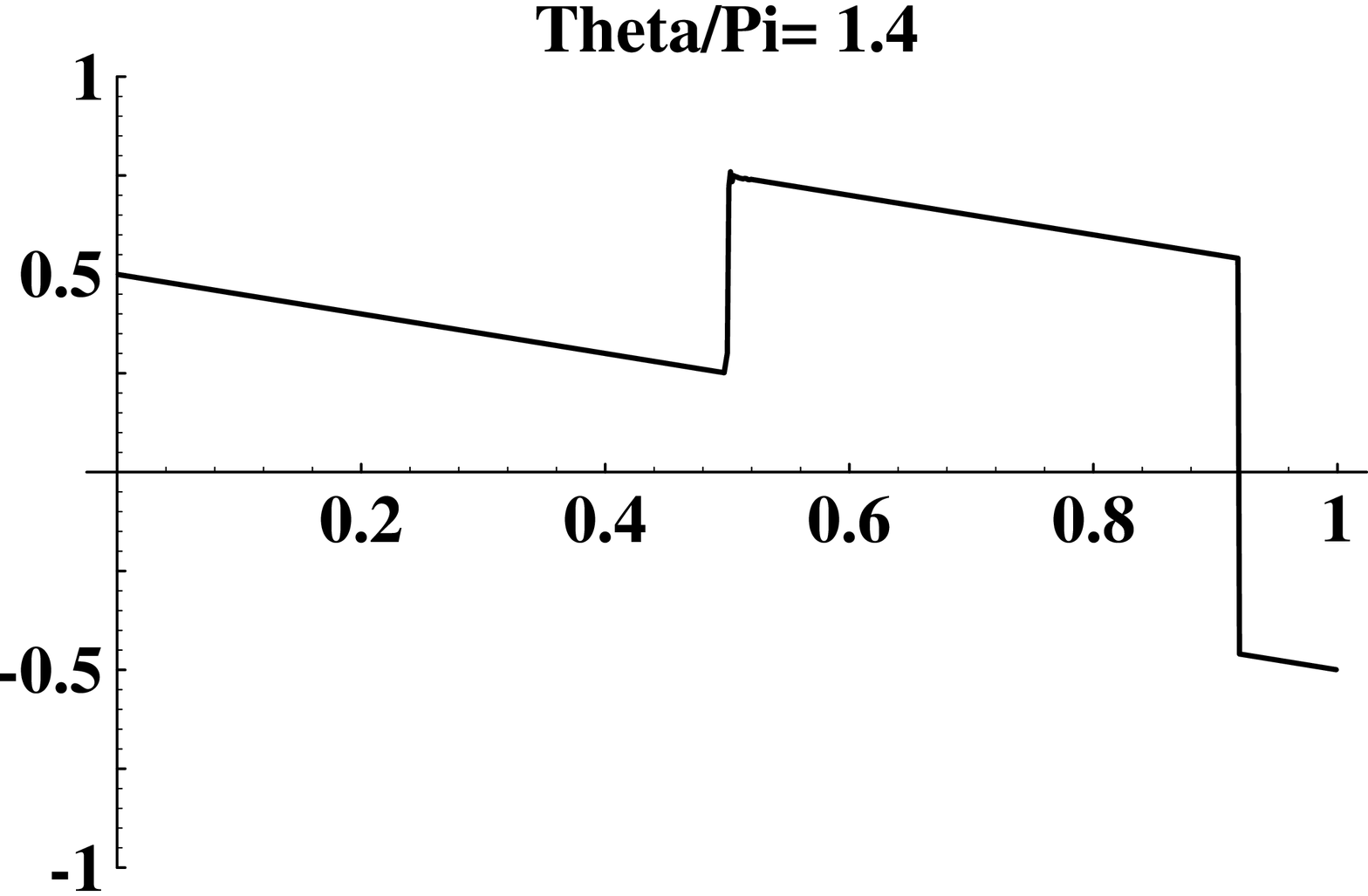}\\%\hline
\end{tabular}
%\end{center}
\vfill
Fig. 3. $\langle N\rangle$ as function of $F$.
\newpage
%\begin{center}
\begin{tabular}{cc}%\hline
\epsfxsize=7cm\epsffile{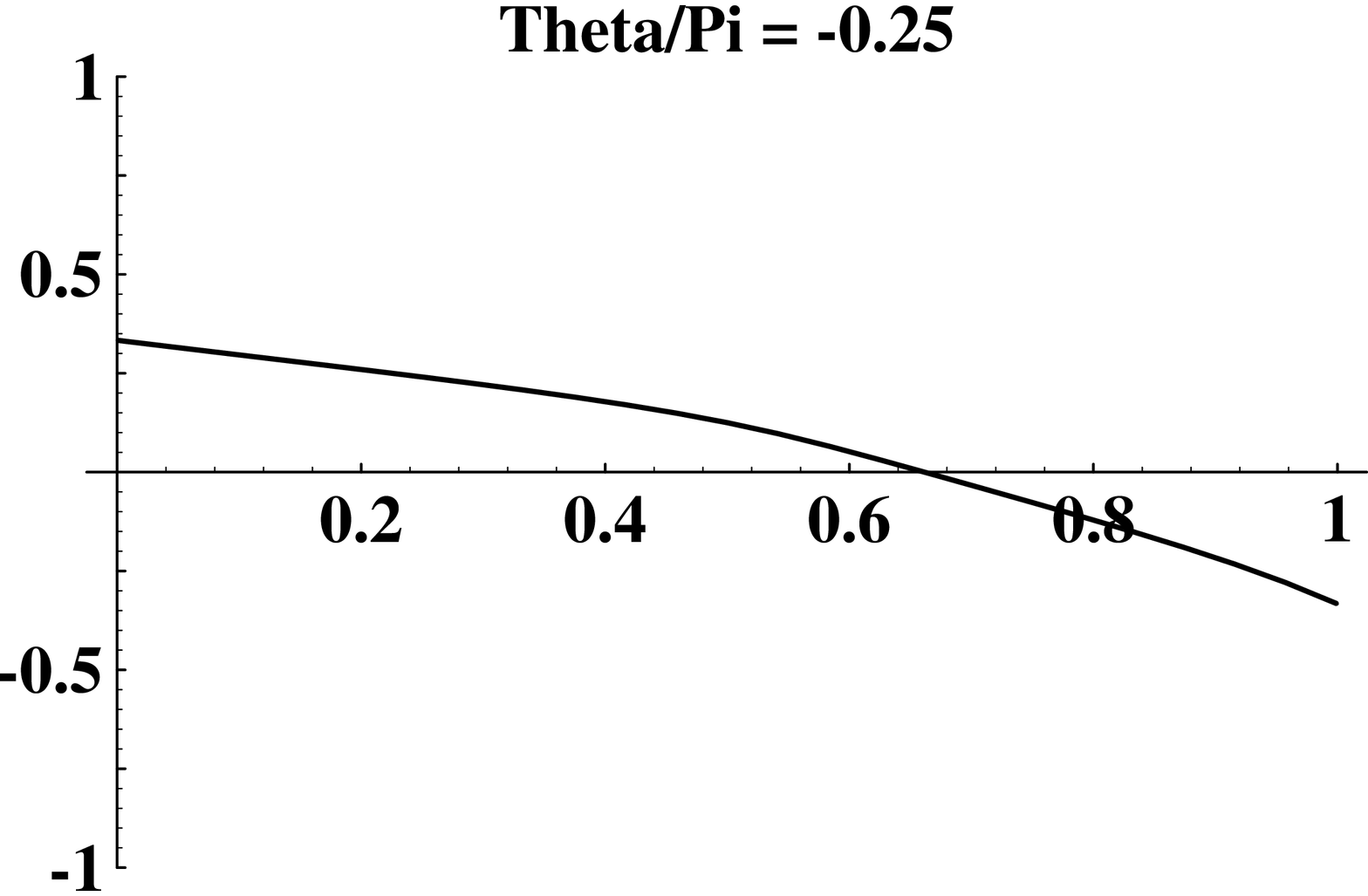}&\epsfxsize=7cm\epsffile{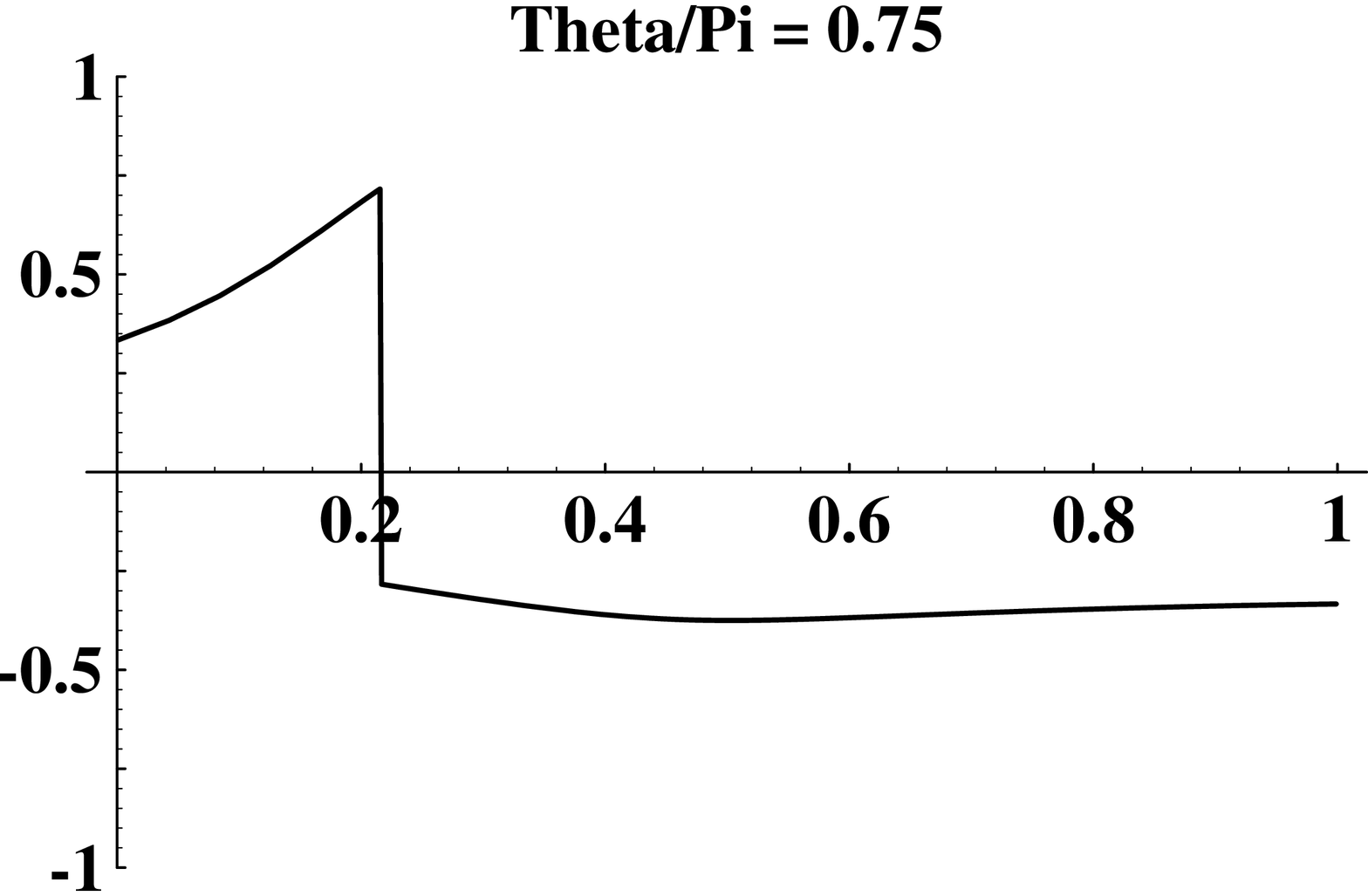}\\%\hline
\epsfxsize=7cm\epsffile{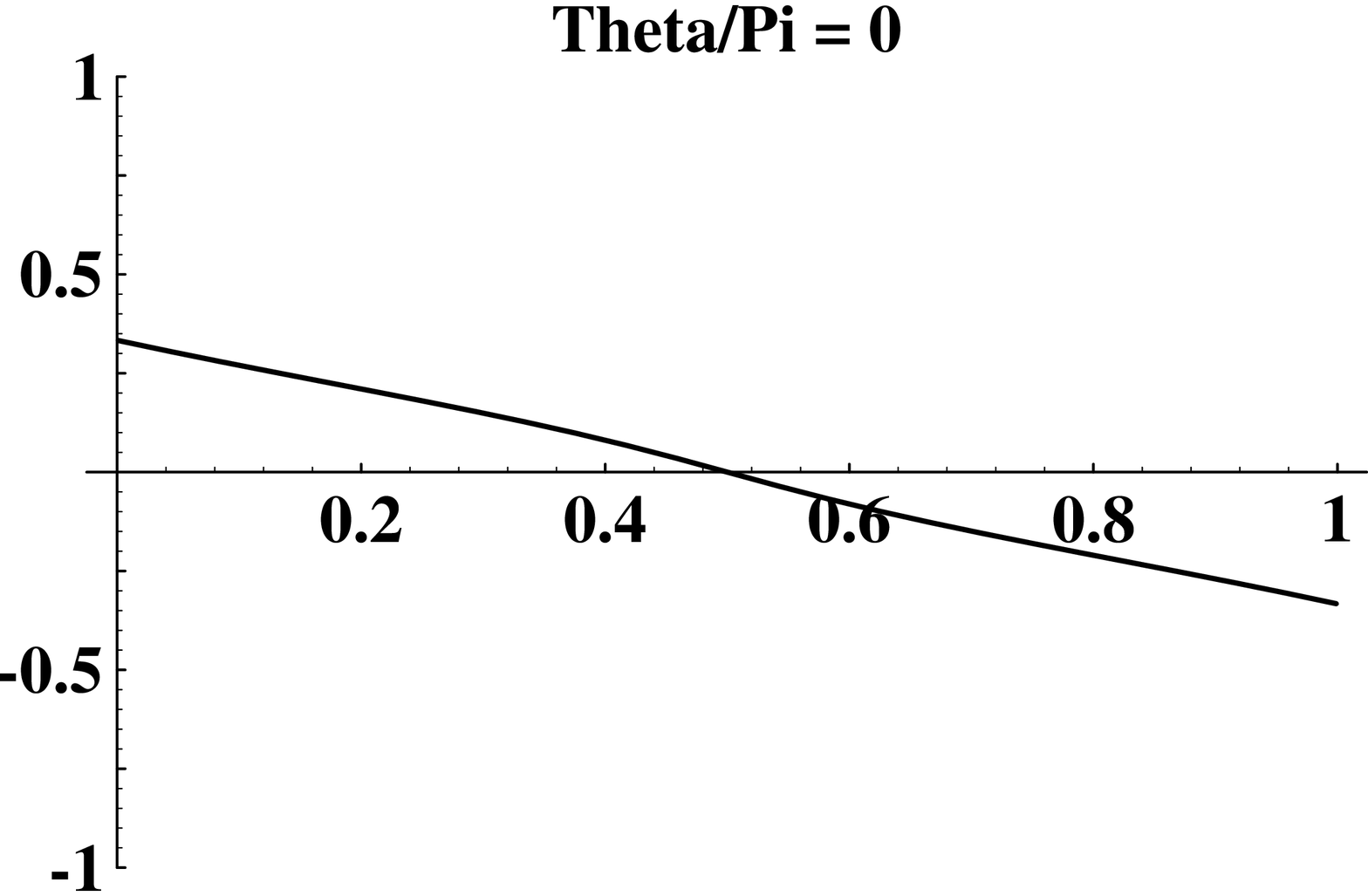}&\epsfxsize=7cm\epsffile{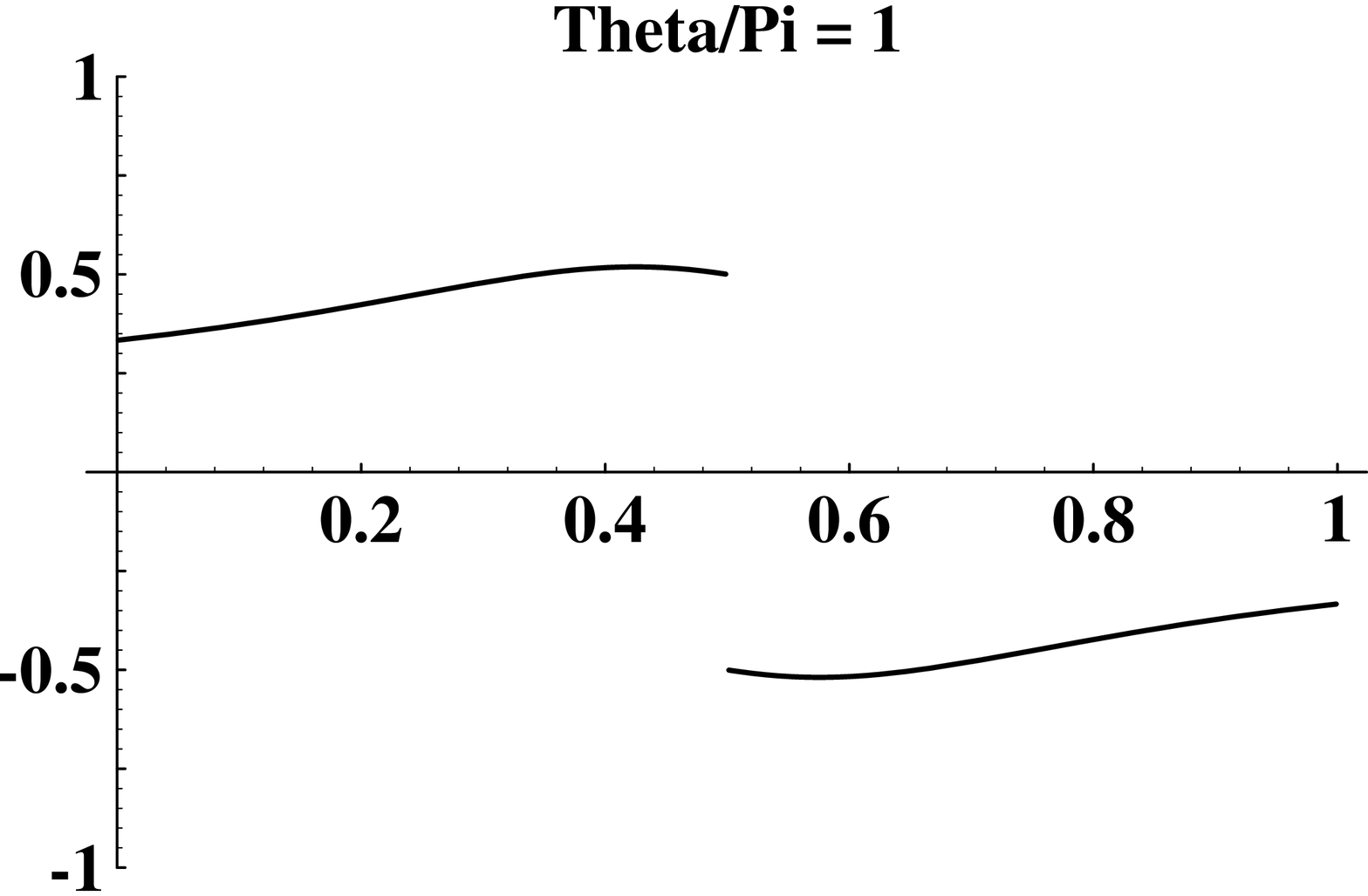}\\%\hline
\epsfxsize=7cm\epsffile{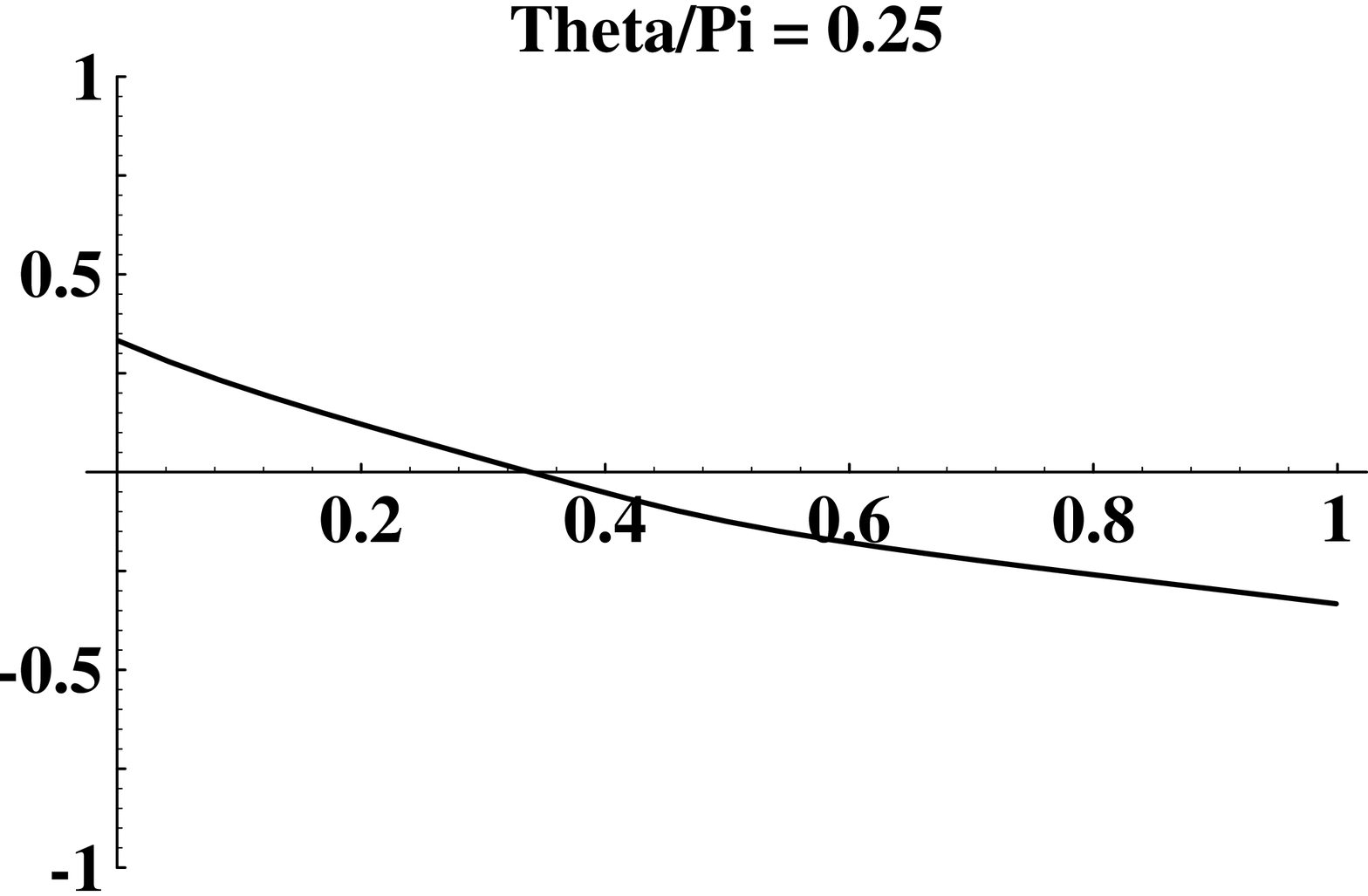}&\epsfxsize=7cm\epsffile{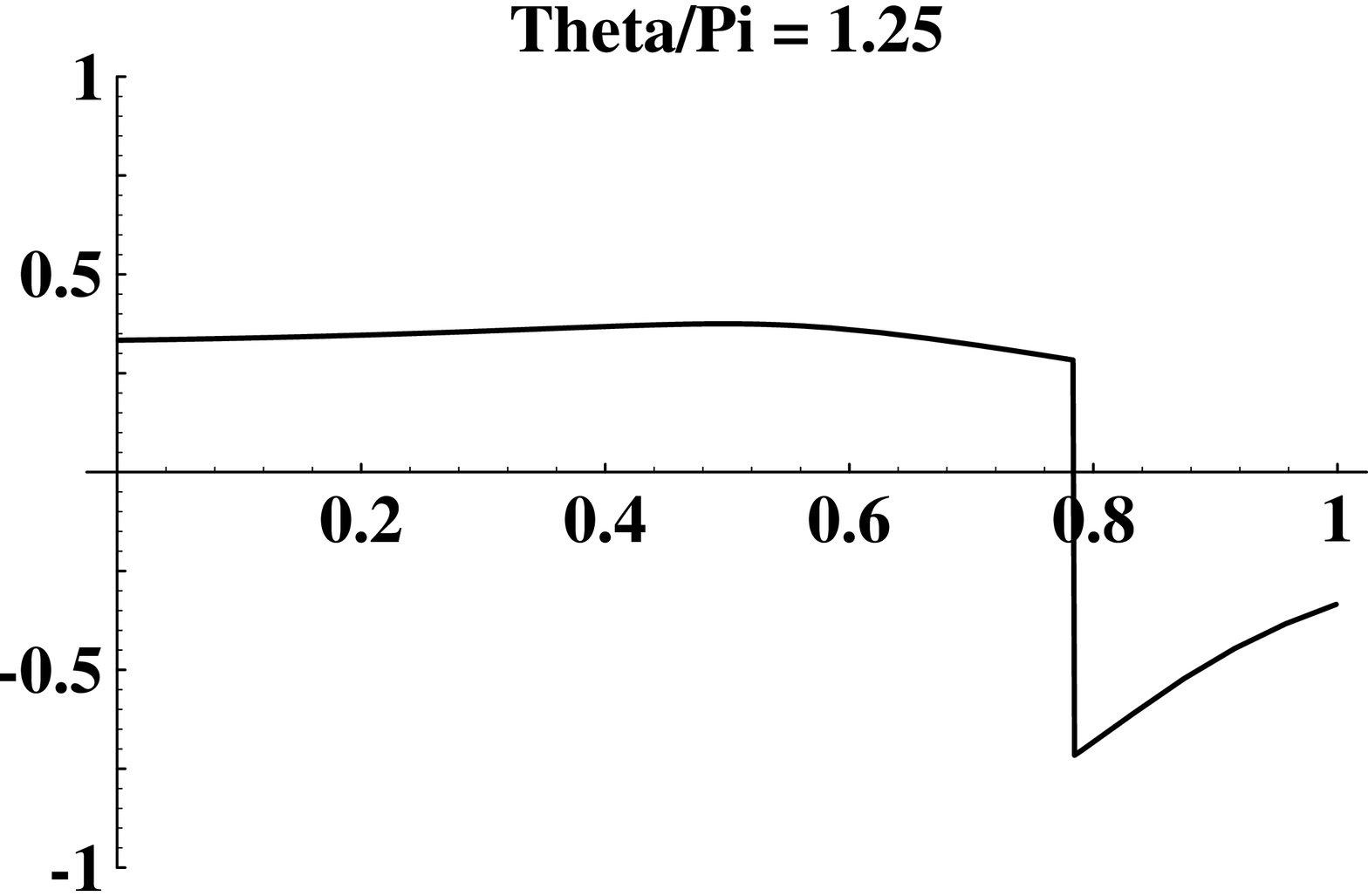}\\%\hline
\end{tabular}
%\end{center}
\vfill
Fig. 4. $2\pi m[e^2F(1-F)]^{-1}\Phi^{(I)}$ as function of $F$.
\end{center}

\end{document}